\documentclass[a4paper,11pt]{article}
\pdfoutput=1 

\usepackage{jheppub} 

\usepackage[T1]{fontenc} 

\def\d{{\mathrm{d}}}

\def\adot{{\dot{\alpha}}}
\def\bdot{{\dot{\beta}}}
\def\gdot{{\dot{\gamma}}}
\def\dotd{{\dot{\delta}}}

\def\btheta{{\bar{\theta}}}
\def\blambda{{\bar{\lambda}}}
\def\bD{{\bar{D}}}

\def\bLambda{{\bar{\Lambda}}}

\def\bu{{\overline{u}}}

\def\ahat{{{\hat{\alpha}}}}
\def\bhat{{{\hat{\beta}}}}
\def\ghat{{{\hat{\gamma}}}}
\def\dhat{{{\hat{\delta}}}}

\def\hlambda{{\widehat{\lambda}}}
\def\hw{{\widehat{w}}}
\def\hN{{\widehat{N}}}
\def\hpsi{{\hat{\psi}}}

\def\tlambda{{\widetilde{\lambda}}}

\def\tw{{\widetilde{w}}}

\def\tD{{\widetilde{D}}}
\def\tbD{{\bar{\widetilde{D}}}}

\def\Ap{{A^\prime}}
\def\Bp{{B^\prime}}

\def\bib{\bibitem}
\def\be{\begin{equation}}
\def\ee{\end{equation}}
\def\ba{\begin{eqnarray}}
\def\ea{\end{eqnarray}}

\usepackage{color}

\title{\boldmath{Open-closed superstring amplitudes using vertex operators in $\mathrm{AdS}_5 \times \mathrm{S}^5$}}

 \author{Thales Azevedo}
 \author{and Nathan Berkovits}
\affiliation{ICTP South American Institute for Fundamental Research\\
Instituto de F\'isica Te\'orica, UNESP -- Univ. Estadual Paulista\\
Rua Dr. Bento T. Ferraz 271, 01140-070, S\~ao Paulo, SP, Brasil}

\emailAdd{thales@ift.unesp.br}
\emailAdd{nberkovi@ift.unesp.br}

\abstract{Using the pure spinor formalism, a particular superstring scattering amplitude involving one closed string and $N$ open string vertex operators in $\mathrm{AdS}{}_5 \times \mathrm{S}^5$ is studied.  It is shown that the tree-level amplitude containing one supergravity state and $N$ super-Yang-Mills states located on D3-branes near the AdS${}_5$ boundary  can be expressed as a $d=4$ ${\cal N}=4$ harmonic superspace integral in terms of the
supergravity and super-Yang-Mills superfields. }

\begin{document} 
\maketitle
\flushbottom

\section{Introduction}

In order to study superstrings in an ${\mathrm{AdS}}_5 \times \mathrm{S}^5$ background \cite{maldacena}, it is possible to use both the Green--Schwarz \cite{metsaev,frolov} and the pure-spinor \cite{nathan,mazzucato} formalisms. Although the superstring action is known in both formalisms, the explicit superfield construction of the vertex operators of the theory is still an open problem. Vertex operators correspond to physical states in string theory, and knowing their expressions is necessary  to compute scattering amplitudes. 


Even though the RNS formalism cannot be used to describe ${\mathrm{AdS}}_5 \times \mathrm{S}^5$, since it is a Ramond-Ramond background,  flat-space RNS vertex operators have been used to compute scattering amplitudes in that background in certain limits \cite{minahan1,minahan2,minahan3}. In the pure-spinor formalism, the first work on vertex operators in ${\mathrm{AdS}}_5 \times \mathrm{S}^5$ was \cite{chandia}. There the authors have constructed massless vertex operators corresponding to on-shell fluctuations around that background, but the expansion in powers of $\theta$ was not computed and the connection between the vertex operators and the
duals of the half-BPS operators was not found.

Recently \cite{fleury}, a  step in that direction was taken. Using the pure-spinor formalism, vertex operator expressions have been found for a particular case, namely the massless states (supergravity)  in the limit $z \to 0$, i.e. $\!\!$close to the AdS boundary ($z$ is the radial coordinate of AdS${}_5$). The vertex operators are in the ghost-number $+2$ cohomology of a BRST operator, and come in a family $\{V_{(N)}\}$ ($N=1,2,\ldots$) such that $V_{(N)}$ is dual to a half-BPS operator involving $N$ super-Yang--Mills fields.

Half-BPS operators can be described in an elegant manner in harmonic superspace as \cite{west,howe,hartwell}
\be
W^{(N)}(x,\theta,u)  := (uu)^{i_1j_1}\ldots(uu)^{i_Nj_N} \,\mathrm{tr}\left[W_{i_1j_1}(x,\theta)\ldots W_{i_Nj_N}(x,\theta)\right],
\label{sohnius}
\ee
where $W_{ij}$ is the ${\mathcal{N}}$ = 4, $d$ = 4 Sohnius superfield strength \cite{sohnius}. In addition, the supergravity vertex operators $V_{(N)}$ in \cite{fleury} were written in terms of harmonic superfields $T^{(4-N)}$ which where shown by Heslop and Howe \cite{howe} to couple naturally to $W^{(N)}$ via
\be
\int\d^4x\int\d u\,\bD^4 D^{\prime 4}\left[W^{(N)}(x,\theta,u) \,T^{(4-N)}(x,\theta,u,\bu)\right].
\label{coupling}
\ee
This led to the following conjecture: the tree-level (disk) scattering amplitude with one closed string supergravity vertex operator $V_{(N)} \propto T^{(4-N)}$
in the bulk and $N$ open string SYM vertex operators located on D3-branes near the AdS${}_5$ boundary would be proportional to the coupling (\ref{coupling}).

In this paper, we will prove this conjecture is indeed true. Using (super)symmetry and BRST arguments, we will show that
\be
{\mathcal{M}}_N   :=   \int \d\xi_1\ldots\d\xi_{N-1}\,{\Big\langle V_{(N)}\,V_\mathrm{SYM}\,U_\mathrm{SYM}(\xi_1)\ldots U_\mathrm{SYM}(\xi_{N-1})\Big\rangle}_{\mathrm{D3-brane}}
\label{onefour}
\ee
can be written as (\ref{coupling}).
Here $V_\mathrm{SYM}$ is the unintegrated  vertex operator of SYM, $U_\mathrm{SYM}$ is the integrated one and  the ``D3-brane'' subscript indicates that these vertex operators are located on D3-branes parallel and close to the AdS${}_5$ boundary. Each of the vertex operators depends on the four transverse D3-brane directions $x^a$ in the plane-wave form $e^{\mathrm{i}k_a^{(r)} x^a(\xi_r)}$. However, when scattered with the closed string state represented by $T^{(4-N)}$ in the limit where the D-branes approach the AdS${}_5$ boundary,  
there are no poles in $k^{(r)}\cdot k^{(s)}$ and the amplitude only depends on $k_a^{(r)}$ through the usual conservation term $\delta^4 (\sum_r k_a^{(r)})$. So it can be expressed as a local integral over $\int \d^4 x$ as in  (\ref{coupling}).

 In principle, the complete expression for the supergravity vertex operators is needed to compute scattering amplitudes, while, as mentioned above, only their leading-order behavior near the boundary is known. However, in the limit where the D-branes approach the AdS${}_5$ boundary, the leading-order behavior is sufficient. 
 
 To understand why the structure of (\ref{onefour}) is so simple, note that $W^{(N)}$ of (\ref{sohnius}) is the half-BPS super-Yang-Mills operator which is dual to the supergravity state represented by $T^{(4-N)}$. For half-BPS states, the duality relation between super-Yang-Mills operators and closed string states is protected against corrections and is given in harmonic superspace by the relation of (\ref{coupling}).

This paper is organized as follows. In section 2, we give a brief review of the computation of the zero-mode cohomology of the ${\mathrm{AdS}}_5 \times \mathrm{S}^5$ BRST operator, obtaining explicit superfield expressions for the behavior of supergravity vertex operators near the boundary of AdS${}_5$. In section 3, we use the pure-spinor prescription and supersymmetry to compute the above-mentioned tree-level amplitude, first in the case $N=1$ and then  extending the result to $N > 1$. Finally, we summarize our results in section 4. The appendix contains notations, conventions and further information useful for the reader.

\section{Review of zero-mode cohomology in $\mathbf{AdS}{}_\mathbf{5} \times \mathbf{S}^\mathbf{5}$}

In this section, we briefly review the computation of the ${\mathrm{AdS}}_5 \times \mathrm{S}^5$ supergravity vertex operators done in \cite{fleury}.

In the pure-spinor formalism, physical vertex operators must be in the cohomology of the BRST operator $Q$. In \cite{fleury}, the authors computed $Q$ for the ${\mathrm{AdS}}_5 \times \mathrm{S}^5$ background using the coset
\be
\frac{\mathrm{PSU}(2, 2|4)}{\mathrm{SO}(4,1) \times \mathrm{SO}(6)} \times \frac{\mathrm{SO}(6)}{\mathrm{SO}(5)}\,,
\label{newcoset}
\ee
instead of the more usual coset $\frac{\mathrm{PSU}(2, 2|4)}{\mathrm{SO}(4,1) \times \mathrm{SO}(5)}$.

The AdS${}_5$ superspace is parameterized by five bosonic variables denoted $z$ and $x^a$ for $a=0$ to 3 and thirty-two fermionic variables denoted $\theta^{\alpha i}, \bar{\theta}^\adot_i, \psi^\beta_j, \bar{\psi}^{\bdot j}$ for $\alpha,\adot=1$ to 2 and $i,j=1$ to 4. We use the standard d=4 two-component spinor notation as described in Appendix \ref{twocomp}. These variables appear in the $\frac{\mathrm{PSU}(2, 2|4)}{\mathrm{SO}(4,1) \times \mathrm{SO}(6)}$ coset representative as
\be
g = \exp \left(x^a P_a + \mathrm{i}\,\theta^{\alpha i} q_{\alpha i} + \mathrm{i}\,\bar{\theta}_{\adot i}\bar{q}^{\adot i}\right) \; \exp \left(\mathrm{i}\,\psi^\beta_j s^j_\beta + \mathrm{i}\,\bar{\psi}^j_\bdot \bar{s}^\bdot_j\right) \; z^\Delta\,,
\ee
where $P_a, q_{\alpha i}, \bar{q}^{\adot i}$ are generators of the $\mathcal{N} = 4$, $d = 4$ supersymmetry algebra, $s_\beta^j, \bar{s}^\bdot_j$ are the $\mathcal{N} = 4$, $d = 4$ superconformal generators and $\Delta$ is the generator of dilatations. With this choice of coset representative, the boundary of AdS${}_5$ is located at $z=0$ and, at the boundary, the variables $x^a, \theta^{\alpha i}, \bar{\theta}^\adot_i$ transform in the usual $\mathcal{N} = 4$, $d = 4$ superconformal manner under the action of global $\mathrm{PSU}(2, 2|4)$ transformations.

The S${}^5$ is parameterized by an SO(6) vector $y^I$ for $I=1$ to 6 satisfying $y^Iy^I=1$. The SO(6) Pauli matrices $(\rho_I)_{ij}$ described in Appendix \ref{dimred} can be used to define $y_{ij} := (\rho_I)_{ij}\,y^I$, which satisfies $y_{ij} = \frac{1}{2}\varepsilon_{ijk\ell}\,y^{k\ell}$ and $y_{ij}y^{jk}=\delta_i^k$.

The final ingredients needed to study the pure-spinor string in ${\mathrm{AdS}}_5 \times \mathrm{S}^5$ are the left- and right-moving ghost variables $\lambda, \hlambda$ and their conjugate momenta $w, \hw$. The $\lambda$'s are pure spinors, i.e. $\!$they satisfy $\lambda \gamma^\mu \lambda = 0$ and $\hlambda \gamma^\mu \hlambda = 0$ for $\mu=0$ to 9. Note these expressions have been written in ten-dimensional notation  where $\lambda^\ahat$ and $\hlambda^\ahat$ are chiral spinors ($\ahat = 1$ to 16) which can be
decomposed into $\mathrm{SO}(3,1) \times \mathrm{SO}(6)$ spinors $(\lambda^{\alpha i},
\blambda^\adot_j)$ and $(\hlambda^{\alpha i},
\bar{\hlambda}{}^\adot_j)$ in the usual manner.
The gauge invariance under $\delta w = (\gamma^\mu \lambda)\,\Lambda_\mu$ for any $\Lambda_\mu$ implies that $w$ can only appear in combinations of the gauge-invariant quantities
\be
N^{\mu\nu}  = \frac{1}{2} w \gamma^{\mu\nu} \lambda\,, \qquad J_\mathrm{g} = w\lambda\,,
\ee
which are respectively the SO($9,1$) Lorentz currents in the ghost sector and the ghost-number current.
Of course, similar expressions hold for the hatted quantities.

\subsection{Worldsheet action}

To construct the BRST-invariant worldsheet action using the coset (\ref{newcoset}), we need to define the left-invariant current $J = g^{-1}\d g$, taking values in the $\mathrm{PSU}(2, 2|4)$ Lie algebra. Here $\d = \d\zeta\frac{\partial}{\partial\zeta} + \d\bar{\zeta}\frac{\partial}{\partial{\bar{\zeta}}}$ and the variables $\zeta, \bar{\zeta}$ parameterize the string worldsheet. The components of this current are defined via
\be
g^{-1}\frac{\partial}{\partial\zeta} g = J^a\frac{1}{2}(P_a + K_a) + J^\Delta\Delta + J^k_jR^j_k + J^{\alpha i}q_{\alpha i} + J_{\adot i}\bar{q}^{\adot i} + J^\beta_js^j_\beta + J^j_\bdot\bar{s}_j^\bdot + \cdots\,,
\ee
where $K_a$ are the generators of special conformal transformations in four dimensions, $R^j_k$ are the SU(4) R-symmetry generators and the dots stand for terms proportional to generators in the isotropy group of AdS${}_5$, i.e.  $\!\!$in SO($4,1$). Analogously, one can define $\bar{J}^a, \bar{J}^\Delta, \ldots$ through the calculation of $g^{-1}\frac{\partial}{\partial{\bar{\zeta}}}g$.

In terms of these currents, the matter part of the worldsheet action is given by
\ba
S_{\mathrm{matter}} & = & \int \d^2\zeta\,\biggr[\frac{1}{2}\eta_{ab}J^a\bar{J}^b + \frac{1}{2}J^\Delta\bar{J}^\Delta -\frac{1}{8}(\nabla y)_{jk}(\bar{\nabla} y)^{jk} \nonumber \\
& & \qquad \qquad -\,2 J^{\alpha i} \bar{J}_{\alpha i} - 2 J_{\adot i}\bar{J}^{\adot i} - 2 J^\beta_j\bar{J}_\beta^j - 2 J_\bdot^j\bar{J}_j^\bdot \\ \nonumber
& &   \qquad \qquad     -\,y_{jk} J^{\alpha j}\bar{J}_\alpha^k - y^{jk}J^\alpha_j \bar{J}_{\alpha k} + y^{jk}J_{\adot j}\bar{J}^\adot_k + y_{jk}J^j_\adot\bar{J}^{\adot k} \biggr]\,,
\ea
where $(\nabla y)_{jk}= \partial y_{jk} + 2J^\ell_{[j}y_{k]\ell}$, ${\partial} := \partial/\partial\zeta$.  One can show that this action is in agreement with the one written in terms of the usual coset (\ref{usualcoset}) by using the SO(6) gauge invariance to gauge-fix $y^I = \delta^I_6$ in $S_\mathrm{matter}$ and then comparing with the well-known result in \cite{mazzucato}, for example.

The complete action has also a contribution coming from the ghosts:
\ba
S_{\mathrm{ghost}} & = & \int \d^2\zeta\,\biggr[ w\cdot \bar{\nabla}\lambda - \hw\cdot \nabla\hlambda  + \frac{1}{2}y^{j\ell}(\bar{\nabla} y)_{\ell k} w_{\mathcal{A}j}\lambda^{\mathcal{A}k} - \frac{1}{2} y^{j\ell}(\bar{\nabla}y)_{\ell k}\hw_{\mathcal{A}j}\hlambda^{\mathcal{A}k} \nonumber \\
& & \qquad \qquad -\, 2N_{ab}\hN^{ab} - 4 (y^J N_{Ja})(y_K \hN^{Ka}) + 2 N_{IJ}\hN^{IJ} - 4 (y^LN_{LJ})(y_M\hN^{MJ})\biggr]\,,\nonumber\\
\ea
where $\lambda^{\mathcal{A}k}, w_{\mathcal{A}j}$ and their hatted counterparts are $\mathrm{SO}(4, 1) \times \mathrm{SO}(6)$ spinors given by
\be
\lambda^{\mathcal{A}k} = \left(\begin{array}{c} \lambda_\alpha^k \\ y^{k\ell}\blambda^\adot_\ell \end{array}\right), \qquad
w_{\mathcal{A}j} = \left(\begin{array}{c} w_j^\alpha \\ -y_{j\ell}\bar{w}_\adot^\ell \end{array}\right),
\ee
and
\ba
w\cdot \bar{\nabla}\lambda & := & w_j^\alpha\bar{\partial}\lambda^j_\alpha + \bar{w}_\adot^k y_{ki} \bar{\partial}(y^{ij}\blambda_j^\adot) - w_j^\alpha \bar{J}_\alpha^\beta\lambda^j_\beta - \bar{w}_\adot^k \bar{J}^\adot_\bdot \blambda_k^\bdot\nonumber\\
& &   +\;2 w_j^\alpha \bar{J}_{\alpha\adot}y^{jk}\blambda_k^\adot - 2\bar{w}_\adot^k y_{k\ell} \bar{J}^{\adot\alpha}\lambda_\alpha^\ell + w_j^\alpha\bar{J}^j_k\lambda^k_\alpha + \bar{w}^k_\adot y_{ki}\bar{J}^i_my^{mn}\blambda_n^\adot\,,
\ea
where $\bar{\partial} := \partial/\partial\bar{\zeta}$. Note that the covariant derivative above has contributions coming from all the $\mathrm{SO}(4, 1) \times \mathrm{SO}(6)$ connections, corresponding to $R^j_k$, $\frac{1}{2}(P_a - K_a)$ and the four-dimensional Lorentz generators $M_{ab}$. An analogous definiton holds for $\hw\cdot \nabla\hlambda$.

\subsection{The BRST operator and its cohomology}\label{BRSTcohomology}

In the last subsection we introduced the worldsheet action for the pure-spinor superstring in an ${\mathrm{AdS}}_5 \times \mathrm{S}^5$ background, which is  given by $S = S_\mathrm{matter} + S_\mathrm{ghost}$. This action is invariant under the BRST transformations generated by
\ba
Q & = & \int \d\zeta\,\Big[\lambda^{\alpha i}(J_{\alpha i} - y_{ij}J^j_\alpha) - \blambda_{\adot i}(J^{\adot i} + y^{ij}J_j^\adot)\Big]\nonumber\\
& - & \int\d\bar{\zeta}\,\Big[\hlambda^{\alpha i}(\bar{J}_{\alpha i} + y_{ij}\bar{J}^j_\alpha) + \bar{\hlambda}_{\adot i}(\bar{J}^{\adot i} - y^{ij}\bar{J}_j^\adot)\Big]\,.
\ea

To simplify the analysis of the cohomology, it is convenient to express the BRST charge in terms of the worldsheet variables and their canonical momenta, defined as $P_x = \frac{\delta S}{\delta(\partial_\tau x)}$, $P_z = \frac{\delta S}{\delta(\partial_\tau z)}$ and so on. Here $\tau$ is the variable associated with the time direction of the worldsheet, whereas $\sigma$ is associated with the space direction. In our conventions,
\be
\partial = \frac{1}{2}(\partial_\sigma - \partial_\tau)\,, \qquad \bar{\partial} = \frac{1}{2}(\partial_\sigma + \partial_\tau)\,.
\ee

After substituting the currents, the operator $Q$ can be organized in the form
\be
Q = Q_{-\frac{1}{2}} + Q_{\frac{1}{2}} + Q_{\frac{3}{2}} + \cdots\,,
\label{expansaoQ}
\ee
where $Q_n \propto z^n$. Near the boundary of AdS${}_5$, it is also possible to expand a physical vertex operator $V$ as
\be
V = V_{d_0} + V_{d_0+1} + \cdots\,,
\label{expansaoV}
\ee
where $V_n \propto z^n$ and $d_0$ is the minimum degree of $V$, i.e. $\!V_{d_0}$ is the leading term in the expansion of $V$ near the boundary.

Equations (\ref{expansaoQ}) and (\ref{expansaoV}) imply that, in order to compute the cohomology of $Q$, one can first compute the cohomology of $Q_{-\frac{1}{2}}$, then compute the cohomology of $Q_{\frac{1}{2}}$ restricted to that of $Q_{-\frac{1}{2}}$, and so on. The reason is that, collecting the terms with the same power of $z$, one has
\be
QV = 0 \;\Longleftrightarrow \; \left\{\begin{array}{l} Q_{-\frac{1}{2}}V_{d_0}=0\,,\\
Q_{\frac{1}{2}}V_{d_0} + Q_{-\frac{1}{2}}V_{d_0+1}=0\,,\\
\vdots
\end{array}
\right.
\label{Q-closed}
\ee
This procedure is well defined since the complete BRST operator $Q$ is nilpotent, which implies $\{Q_{\frac{1}{2}},Q_{\frac{1}{2}}\} + 2\,\{Q_{-\frac{1}{2}},Q_{\frac{3}{2}}\} = 0$, that is, $Q_{\frac{1}{2}}$ is nilpotent when acting on states in the cohomology of $Q_{-\frac{1}{2}}$. The same nilpotency argument applies for $Q_{\frac{3}{2}}, Q_{\frac{5}{2}}$, and so on.

The computation of $Q_{-\frac{1}{2}}$ gives
\be
Q_{-\frac{1}{2}} \propto \frac{1}{\sqrt{z}}\left(\lambda^{+ \gamma m} y_{mi} P_{\psi_i^\gamma} + \blambda^{+\adot}_j y^{ji} P_{\bar{\psi}^{i\adot}}\right) + \partial_\sigma\mathrm{-terms}\,,
\ee
where
\be
\lambda^{+\alpha i} := -\mathrm{i}\,(\lambda^{\alpha i} - \hlambda^{\alpha i})\,, \qquad \blambda^+_{\adot j} := \mathrm{i}\,(\blambda_{\adot j} + \bar{\hlambda}_{\adot j})\,.
\ee
We do not consider terms in $Q$ containing $\sigma$-derivatives, since  we are only interested in its zero-mode cohomology. In other words, we take the limit in which the string length goes to zero (or the tension goes to infinity), which corresponds to supergravity (massless states). Because of the usual quartet argument, we assume the zero-mode cohomology of $Q_{-\frac{1}{2}}$ is independent of $\lambda^+$.\footnote{Actually, this argument is too naive, and the cohomology is only independent of $\lambda^+$ after allowing dependence on non-mininal pure-spinor variables. See fotenote number 3 of \cite{fleury} for a more detailed discussion.} Moreover, the states in the cohomology depend on $\psi$ only through $\lambda^-\gamma_\mu \hat{\psi}$, where $\hat{\psi} := y_J\,(\gamma^J\psi)$ and
\be
\lambda^{-\alpha i} := -\mathrm{i}\,(\lambda^{\alpha i} + \hlambda^{\alpha i})\,, \qquad \blambda^-_{\adot j} := \mathrm{i}\,(\blambda_{\adot j} - \bar{\hlambda}_{\adot j})\,.
\ee

It is easy to see that $\lambda^-\gamma_\mu \hat{\psi}$ is annihilated by $Q_{-\frac{1}{2}}$, since $Q_{-\frac{1}{2}}\left(\lambda^-\gamma_\mu \hat{\psi}\right) \propto \lambda^-\gamma_\mu\lambda^+$, which vanishes because of the pure-spinor conditions for $\lambda$ and $\hlambda$. One can also show that
\be
\lambda^+ \gamma_\mu \lambda^+ + \lambda^- \gamma_\mu \lambda^- = 0\,.
\ee
This identity implies that, when considering states in the cohomology of $Q_{-\frac{1}{2}}$, we have $\lambda^-\gamma_\mu\lambda^- = 0$, i.e. $\!\lambda^-$ is a pure spinor.

The next step is to compute the zero-mode cohomology of $Q_{\frac{1}{2}} + Q_{\frac{3}{2}} + \cdots$ restricted to states in the cohomology of $Q_{-\frac{1}{2}}$. This means we can neglect terms containing $\lambda^+$ and we can consider $\lambda^-$ a pure spinor. It turns out that $Q_{\frac{3}{2}}, Q_{\frac{5}{2}}, \ldots$ act as zero on states in the cohomology of $Q_{-\frac{1}{2}}$. This is because the terms depending on ${\psi}$ in their expansions cannot be expressed in terms of the $\lambda^-\gamma_\mu\hat{\psi}$. Thus, the zero-mode cohomology of $Q$ near the boundary of AdS${}_5$ is determined by $Q_{-\frac{1}{2}}$ and $Q_{\frac{1}{2}}$ only.

Writing the canonical momenta as derivatives, the part of $Q_{\frac{1}{2}}$ that is relevant for us is, then,
\be
Q_{\frac{1}{2}} \propto \sqrt{z}\left[\lambda^\ahat D_\ahat + 4\,(\lambda \gamma^{[ij]}\hat{\psi})\frac{\partial}{\partial y^{ij}} + y_{ij} (\lambda \gamma^{[ij]}\hat{\psi})\left(2z\frac{\partial}{\partial z} + y^{k\ell}\frac{\partial}{\partial y^{k\ell}} - \lambda^\ahat\frac{\partial}{\partial \lambda^\ahat}             \right)\right],
\label{Qhalf}
\ee
where $D_\ahat = \frac{\partial}{\partial\theta^\ahat} + (\theta\gamma^a)_\ahat \partial_a$ is the dimensional reduction of the d=10 supersymmetric derivative and we have dropped the minus superscript from $\lambda^-$. The second term in (\ref{Qhalf}) is understood not to act on $\lambda\gamma_\mu\hat{\psi}$, even though $\hat{\psi}$ depends on $y$.

In order to express the vertex operators in a convenient way using harmonic superspace, we need to introduce non-minimal pure-spinor variables \cite{nonminimal}. They consist of a bosonic spinor $\tlambda_\ahat$ and a fermionic spinor $r_\ahat$, as well as their conjugates $\tw^\ahat$ and $s^\ahat$. These variables satisfy the constraints
\be
\tlambda \gamma^\mu \tlambda = 0\,, \qquad \tlambda \gamma^\mu r = 0\,.
\ee
The first of these equations implies $\tlambda_\ahat$ is a pure spinor.

After introducing the non-minimal variables, we need to modify the BRST operator $Q_{\frac{1}{2}}$ as:
\be
Q_{\frac{1}{2}} \longmapsto Q_{\frac{1}{2}} + \tw^\ahat r_\ahat\,.
\ee
The addition of this term implies, using the quartet argument, that the cohomology is independent
of the non-minimal variables.

Then, for arbitrary $N$, it was shown in \cite{fleury} that $Q_{\frac{1}{2}} + \tw^\ahat r_\ahat$ annihilates the following vertex operator:
\be
V_{(N)} = z^{2-N} \int\d u \sum_{n=0}^4 8^n P_n(N)\,(yuu)^{N-n-1}\,\Omega_{(n)}T^{(4-N)}(x,\theta,u,\bu)\,,
\label{vertex}
\ee
where $P_n(N) = \frac{1}{N}\prod_{m=0}^n(N-m)$ is a polynomial of degree $n$ in $N$, $T^{(4-N)}$ is a G-analytic superfield of harmonic U(1) charge $4-N$, and $\int\d u$ denotes an integral over the compact space SU(4)/S(U(2)$\times$U(2)) parameterized by the harmonic variables $u_\mathcal{I}{}^i$. See appendix \ref{harmspace} for more information on harmonic superspace. The operators $\Omega_{(n)}$ in (\ref{vertex}) are defined by
\begin{subequations}\label{omega}
\ba
\Omega_{(0)}  & = &  \frac{1}{16}(uu)^{ij}(\lambda\tlambda)^{-2}(\tlambda\gamma_{\mu\nu\rho\sigma[ij]}\tlambda)(\lambda\gamma^\mu\tD)(\lambda\gamma^\nu\tD)(\lambda\gamma^\rho\tD)(\lambda\gamma^\sigma\tD) \nonumber\\
 & &  +\;\frac{1}{\sqrt{z}}(\lambda\tlambda)^{-2}(r\gamma_{\mu\nu\rho}\tlambda)(\lambda\gamma^\mu\tD)(\lambda\gamma^\nu\tD)(\lambda\gamma^\rho\tD)\,, \\
\Omega_{(1)}  & = &  -\frac{1}{2}(uu)^{ij}(\lambda\tlambda)^{-2}(\tlambda\gamma_{\mu\nu\rho\sigma[ij]}\tlambda)(\lambda\gamma^\mu\hpsi)(\lambda\gamma^\nu\tD)(\lambda\gamma^\rho\tD)(\lambda\gamma^\sigma\tD) \nonumber\\
& & -\;\frac{6}{\sqrt{z}}(\lambda\tlambda)^{-2}(r\gamma_{\mu\nu\rho}\tlambda)(\lambda\gamma^\mu\hpsi)(\lambda\gamma^\nu\tD)(\lambda\gamma^\rho\tD)\,, \\
\Omega_{(2)}  & = &  \frac{3}{2}(uu)^{ij}(\lambda\tlambda)^{-2}(\tlambda\gamma_{\mu\nu\rho\sigma[ij]}\tlambda)(\lambda\gamma^\mu\hpsi)(\lambda\gamma^\nu\hpsi)(\lambda\gamma^\rho\tD)(\lambda\gamma^\sigma\tD) \nonumber\\
& & +\;\frac{12}{\sqrt{z}}(\lambda\tlambda)^{-2}(r\gamma_{\mu\nu\rho}\tlambda)(\lambda\gamma^\mu\hpsi)(\lambda\gamma^\nu\hpsi)(\lambda\gamma^\rho\tD)\,, \\
\Omega_{(3)}  & = &  -2\,(uu)^{ij}(\lambda\tlambda)^{-2}(\tlambda\gamma_{\mu\nu\rho\sigma[ij]}\tlambda)(\lambda\gamma^\mu\hpsi)(\lambda\gamma^\nu\hpsi)(\lambda\gamma^\rho\hpsi)(\lambda\gamma^\sigma\tD) \nonumber\\
& & -\;\frac{8}{\sqrt{z}}(\lambda\tlambda)^{-2}(r\gamma_{\mu\nu\rho}\tlambda)(\lambda\gamma^\mu\hpsi)(\lambda\gamma^\nu\hpsi)(\lambda\gamma^\rho\hpsi)\,, \\
\Omega_{(4)}  & = &  (uu)^{ij}(\lambda\tlambda)^{-2}(\tlambda\gamma_{\mu\nu\rho\sigma[ij]}\tlambda)(\lambda\gamma^\mu\hpsi)(\lambda\gamma^\nu\hpsi)(\lambda\gamma^\rho\hpsi)(\lambda\gamma^\sigma\hpsi)\,, 
\ea
\end{subequations}
with $\widetilde{D}^\ahat := -\frac{1}{2}(\bu\bu)_{ij}(\gamma^{[ij]}D)^\ahat$.

It is easy to check that the $\left(2z\frac{\partial}{\partial z} + y^{k\ell}\frac{\partial}{\partial y^{k\ell}} - \lambda^\ahat\frac{\partial}{\partial \lambda^\ahat}             \right)$-part of $Q_{\frac{1}{2}}$ annihilates $V_{(N)}$. Moreover, the $\Omega_{(n)}$'s have been designed so that the following equations are satisfied:
\begin{subequations}\label{omegahat}
\ba
\left(\sqrt{z} \lambda^\ahat D_\ahat + \tilde{w}^\ahat r_\ahat\right)\Omega_{(0)}T^{(4-N)} & = & 0\,, \label{omegahat.a}\\
\left(\sqrt{z} \lambda^\ahat D_\ahat + \tilde{w}^\ahat r_\ahat\right)\Omega_{(1)}T^{(4-N)} & = & -\sqrt{z}\, (\lambda\gamma^{[ij]}\hat{\psi})(uu)_{ij}\Omega_{(0)}T^{(4-N)}\,, \label{omegahat.b}\\
\left(\sqrt{z} \lambda^\ahat D_\ahat + \tilde{w}^\ahat r_\ahat\right)\Omega_{(2)}T^{(4-N)} & = & -\sqrt{z}\, (\lambda\gamma^{[ij]}\hat{\psi})(uu)_{ij}\Omega_{(1)}T^{(4-N)}\,, \label{omegahat.c}\\
\left(\sqrt{z} \lambda^\ahat D_\ahat + \tilde{w}^\ahat r_\ahat\right)\Omega_{(3)}T^{(4-N)} & = & -\sqrt{z}\, (\lambda\gamma^{[ij]}\hat{\psi})(uu)_{ij}\Omega_{(2)}T^{(4-N)}\,, \label{omegahat.d}\\
\left(\sqrt{z} \lambda^\ahat D_\ahat + \tilde{w}^\ahat r_\ahat\right)\Omega_{(4)}T^{(4-N)} & = & -\sqrt{z}\, (\lambda\gamma^{[ij]}\hat{\psi})(uu)_{ij}\Omega_{(3)}T^{(4-N)}\,, \label{omegahat.e}\\
\sqrt{z}\, (\lambda\gamma^{[ij]}\hat{\psi})(uu)_{ij}\Omega_{(4)}T^{(4-N)} & = & 0\,.
\ea
\end{subequations}

\subsection{\boldmath{Connection to $\mathcal{N} = 4$ SYM}}

If the operators (\ref{vertex}) correspond to supergravity states in $\mathrm{AdS}_5 \times \mathrm{S}^5$, then AdS/CFT predicts they should be related to half-BPS states in $\mathcal{N} = 4$ SYM. Indeed, we can make this relation explicit by making use of the harmonic superspace. Consider the following family of gauge-invariant operators  introduced in \cite{west}:
\be
W^{(N)} := \mathrm{tr}\,W^N\,,\qquad N=1,2,\ldots\,,
\ee
where $W:=(uu)^{ij} W_{ij}$, $W_{ij}$ is the Sohnius field strength of ${{\mathcal{N}}}=4$ SYM \cite{sohnius} and the trace is taken over the gauge group. In components, we have 
\be
W_{ij}(x,\theta,\btheta) = \phi_{ij}(x) -\varepsilon_{ijk\ell}\theta^k\xi^\ell(x) + 2\btheta_{[i}\bar{\xi}_{j]}(x) +\frac{1}{4}\varepsilon_{ijk\ell}\theta^k\sigma^{ab}\theta^\ell f_{ab}(x) - \frac{1}{2}\btheta_{i}\tilde{\sigma}^{ab}\btheta_{j}f_{ab}(x) +\cdots\,,
\ee
where $\phi_{ij}$, $\xi$, $\bar{\xi}$ and $f_{ab}$ are, respectively, the $\mathcal{N} = 4$ SYM scalars, chiral and anti-chiral gluinos and gluon field-strength. It is easy to see $W^{(N)}$ describes a gauge-invariant half-BPS operator constructed from $N$
SYM fields.

For each value of $N$, one can show that $W^{(N)}$ is an analytic superfield. Thus, it is possible to define a superfield dual to $W^{(N)}$ through the coupling
\be
\int\d^4x\int\d u\,\bD^4 D^{\prime 4}\left[W^{(N)}(x,\theta,u) \,U^{(4-N)}(x,\theta,u,\bu)\right],
\label{couple}
\ee
where $U^{(4-N)}$ is a (otherwise unconstrained) G-analytic superfield of harmonic U(1) charge $4-N$. In \cite{howe}, these superfields $U^{(4-N)}$ were shown to be in one-to-one correspondence with the chiral superfields describing type IIB supergravity states in $\mathrm{AdS}_5 \times \mathrm{S}^5$.

It is natural to identify $U^{(4-N)}$ with the $T^{(4-N)}$ of (\ref{vertex}), since they have exactly the same properties. One way to see  it is consistent  is by first noting that the coupling (\ref{couple}) is invariant under the transformation
\be
U^{(N)} \longmapsto U^{(N)} + u_A{}^i\frac{\partial}{\partial \bu_\Ap{}^i}\Xi^{(N-1)\,A}_\Ap\,,
\label{symm}
\ee
with $\Xi^{(N-1)\,A}_\Ap$ some G-analytic superfield of harmonic U(1) charge $N-1$. Then one can show that, when $T^{(4-N)}$ changes according to (\ref{symm}), $V_{(N)}$ changes by a BRST-trivial amount.

Thus, the superfield $T^{(4-N)}$ appearing in the expression for the vertex operator $V_{(N)}$ is dual to the half-BPS operator $W^{(N)}$ in the sense of (\ref{couple}). This in turn implies a correspondence between the supergravity state itself and $W^{(N)}$.
For example, when $T^{(2)}=\theta^{\alpha i}(uu)_{ij}\theta^{\beta j}\theta_\alpha^k(uu)_{k\ell}\theta_\beta^\ell$, one can show $V_{(2)} \propto y_{ij}(\lambda^i\lambda^j)$ (up to BRST-trivial terms). This PSU($2,2|4$) scalar is the zero-momentum dilaton vertex operator, which is dual to the linearized SYM action $\int\d^4 x \int\d u\,\bD^4 W^{(2)}$, as can be seen from (\ref{couple}).

\section{Open-closed amplitudes}

Because of the duality presented at the end of the last section, and also because of symmetry arguments, the  disk scattering amplitude with the supergravity vertex operator $V_{(N)}$ and $N$ massless open superstring (SYM) vertex operators was conjectured in \cite{fleury} to be proportional to the coupling (\ref{couple}). The SYM vertex operators would be located on D3-branes parallel and close to the AdS${}_5$ boundary, at some fixed value of $y^{ij}$  and $z$  near 0. Since the disk has an SL($2,\mathbb{R}$) symmetry which allows us to fix the positions of one open and one closed superstring vertex operator, 
the disk amplitude has the form
\be
{\mathcal{M}}_N:={\left\langle V_{(N)}\,V_\mathrm{SYM}\int\d \xi_1 \cdots \d \xi_{N-1}\,U_\mathrm{SYM}(\xi_1) \cdots U_\mathrm{SYM}(\xi_{N-1})\right\rangle}_{\mathrm{D3-brane}},
\label{newmain}
\ee
where $V_\mathrm{SYM}$ is the unintegrated  vertex operator of SYM and $U_\mathrm{SYM}$ is the integrated one.

The angle brackets in the above equation contain integrations over the $x$, $\lambda$ and $\theta$ zero modes, but they do not contain  integrations over the $z$ and $y^{ij}$ zero modes, since the position of the D3-brane is fixed.  Schematically, one has
\be
{\Big\langle \lambda^\ahat \lambda^\bhat \lambda^\ghat\,f(x,z,y,\theta)\Big\rangle}_{\mathrm{D3-brane}} =   \int \frac{\d^4 x}{z^4} \int \frac{(\d^5\theta)^{\ahat\bhat\ghat}}{z^{-5/2}} f(x,z,y,\theta)\,,
\label{limite}
\ee
where the powers of $z$ ensure the measure is  PSU($2,2|4$)-invariant, since $\d^4 x$ and $\d^5\theta$ have dimension $-4$ and $\frac{5}{2}$, respectively.
More details on the integration of the $\lambda$ and $\theta$ zero modes will be given shortly.

Proving that (\ref{newmain}) is indeed proportional to (\ref{couple})  is the main purpose of this paper, and what we begin to do in the following.

\subsection{\boldmath{The case $N = 1$}}

We now proceed to the computation of the amplitude (\ref{newmain}) for the case $N=1$. This is the simplest case, not only because the amplitude does not involve integrated vertex operators, but also because $V_{(1)}$ is simpler than any other supergravity vertex operator in ${\mathrm{AdS}}_5 \times \mathrm{S}^5$. Indeed, from (\ref{vertex}) we have
\be
V_{(1)}=z\int \d u\,\Omega_{(0)} T^{(3)}(x,\theta,u,\bu)\,,
\label{V1}
\ee
which has no $y$- or $\psi$-dependence. These operators are the duals to SYM ``singleton'' operators, i.e.
the duals to abelian SYM fields.

In fact, the expression for $V_{(1)}$ can be further simplified. By adding the BRST-trivial quantity
\be
\left(Q_{\frac{1}{2}} + \tw^\ahat r_\ahat\right)\left[2\sqrt{z}\int\d u\,\frac{1}{(\lambda\tlambda)}(\lambda\gamma^\mu\widetilde{D})(\lambda\gamma^\nu\widetilde{D})(\tlambda\gamma_{\mu\nu}\widetilde{D})T^{(3)}(x,\theta,u,\bu)\right]
\ee
to $V_{(1)}$, we get an equivalent expression which does not depend on the non-minimal pure spinor variables. It looks the same as (\ref{V1}), but with $\Omega_{(0)}$ replaced by
\be
\widetilde{\Omega}_{(0)}  :=   -\frac{1}{4}(uu)^{ij}(\lambda\gamma^\mu\widetilde{D})(\lambda\gamma^\nu\widetilde{D})(\widetilde{D}\gamma_{\mu\nu[ij]}\widetilde{D})\,.
\label{newomega}
\ee

Based on the conjecture referred to at the beginning of this section, we expect the following relation to hold:
\be
{\mathcal{M}}_1={\Big\langle V_{(1)} \,V_\mathrm{SYM}\Big\rangle}_{\mathrm{D3-brane}} \propto \int \d^4 x \int \d u \,\bar{D}^4 D^{\prime 4} \,\big[W^{(1)}(x,\theta,u)\, T^{(3)}(x,\theta,u,\bu)\big]\,,
\label{main}
\ee
where $V_{\mathrm{SYM}} = \sqrt{z}\,\lambda^\ahat A_\ahat(x,\theta)$ and $A_\ahat$ is the dimensional reduction of the ${\mathcal{N}}=1$, d = 10 SYM superfield, whose properties are reviewed in  Appendix \ref{ap.SYM}. Substituting the expressions for the vertex operators, one can write the amplitude as
\be
\mathcal{M}_1 = \int \d^4 x \int \d u \,\left\langle \lambda^\ahat A_\ahat\,\widetilde{\Omega}_{(0)}T^{(3)}\right\rangle,
\ee
where we used (\ref{V1}) and replaced $\Omega_{(0)}$ by the $\widetilde{\Omega}_{(0)}$ of (\ref{newomega}).

Now the angle brackets denote the integrations of the $\lambda$ and $\theta$ zero modes only. In order to perform these integrations, we need to find a $ \lambda^\ahat D_\ahat $-invariant, $\mathrm{SO}(3,1)\times\mathrm{SU}(4)$ scalar measure factor, since this is the symmetry of the boundary of AdS${}_5\times\mathrm{S}^5$. At first, it might seem to be just a matter of dimensionally reducing
the ten-dimensional measure
\be
(\lambda\gamma^\mu\theta)(\lambda\gamma^\nu\theta)(\lambda\gamma^\rho\theta)(\theta\gamma_{\mu\nu\rho}\theta)\,.
\label{measure}
\ee
However, although this particular combination of $\lambda$'s and $\theta$'s is special in ten flat dimensions, since it is the unique (up to an overall factor) SO($9,1$) scalar which can be built out of three $\lambda$'s and five $\theta$'s, there is no reason why its dimensional reduction should be preferred over any other BRST-invariant, $\mathrm{SO}(3,1)\times\mathrm{SU}(4)$ scalar in four dimensions. Hence, there could be some ambiguity in the definition of the ${\mathcal{N}}=4$, d = 4 measure factor. Fortunately, though, it was shown in \cite{thales} that it is  unique up to BRST-trivial terms and an overall factor, so we can use (\ref{measure}) consistently.

Using the measure factor (\ref{measure}), 
the pure-spinor prescription for the computation of tree-level scattering amplitudes states that \cite{nathan}
\be
\left\langle \lambda^{\ahat_1} \lambda^{\ahat_2} \lambda^{\ahat_3} f_{\ahat_1\ahat_2 \ahat_3}(x,\theta,u,\bu)\right\rangle  \propto   ({\mathcal{T}}D^5)^{\ahat_1\ahat_2 \ahat_3}   f_{\ahat_1\ahat_2 \ahat_3}(x,\theta,u,\bu)\,,
\ee
where
\be
({\mathcal{T}}D^5)^{\ahat_1\ahat_2 \ahat_3}:= {\mathcal{T}}^{\ahat_1\ahat_2 \ahat_3}_{\bhat_1\bhat_2\bhat_3}(\gamma^\mu D)^{\bhat_1} (\gamma^\nu D)^{\bhat_2} (\gamma^\rho D)^{\bhat_3}(D\gamma_{\mu\nu\rho}D)\,,
\label{TD5}
\ee

\be
{\mathcal{T}}^{\ahat_1\ahat_2 \ahat_3}_{\bhat_1\bhat_2\bhat_3} :=\delta^{(\ahat_1}_{\bhat_1} \delta^{\ahat_2}_{\bhat_2} \delta^{\ahat_3)}_{\bhat_3} + \frac{3}{20} (\gamma_\mu)^{(\ahat_1 \ahat_2} \delta_{(\bhat_1}^{\ahat_3)} (\gamma^\mu)_{\bhat_2 \bhat_3)}\,.
\ee
This definition ensures that $({\mathcal{T}}D^5)^{\ahat_1\ahat_2 \ahat_3}$ is totally symmetric and $\gamma$-traceless, i.e.
\be
(\gamma^\mu)_{\ahat_1\ahat_2}({\mathcal{T}}D^5)^{\ahat_1\ahat_2 \ahat_3}=0\,,
\ee
as is the product of three pure spinors.

In the case at hand, 
\be
f_{\ahat\bhat\ghat}(x,\theta,u,\bu) \propto A_\ghat(x,\theta)\, \Omega_{(0)\ahat\bhat}T^{(3)}(x,\theta,u,\bu)\,,
\ee
with
\be
\Omega_{(0)\ahat\bhat} :=  (uu)^{ij}(\gamma^\mu\widetilde{D})_\ahat(\gamma^\nu\widetilde{D})_\bhat (\widetilde{D}\gamma_{\mu\nu[ij]}\widetilde{D})\,.
\label{omega0}
\ee
Thus, we arrive at the final form of the amplitude:
\be
\mathcal{M}_1 \propto \int \d^4 x \int \d u \,({\mathcal{T}}D^5)^{\ahat\bhat\ghat}\left[A_\ghat(x,\theta)\, \Omega_{(0)\ahat\bhat}T^{(3)}(x,\theta,u,\bu)\right].
\label{amplitude}
\ee

\subsubsection{Possible contributions}

Computing (\ref{amplitude}) explicitly would be very complicated, but fortunately we  need not do that. Instead, we can use symmetry arguments and equations of motion to determine the form of the terms that might appear in the computation, and then use supersymmetry to obtain the relative coefficients.

Let us see what kind of terms one can expect to find when computing $\mathcal{M}_1$. Recalling that
\be
\widetilde{D}^\ahat := -\frac{1}{2}(\bu\bu)_{k\ell}(\gamma^{[k\ell]}D)^\ahat = \left(\begin{array}{c}\widetilde{D}^{\alpha i} \\ \tbD{}^\adot_j\end{array}\right) = \left(\begin{array}{c}(\bu\bu)^{ik}D^\alpha_k \\ (\bu\bu)_{j\ell}\bD^{\adot \ell}\end{array}\right),
\ee
we see that the amplitude has the schematic form
\be
{{\mathcal{M}}}_1 \sim \int \d^4 x \int\d u\, \bu^6 (D_\ahat)^5 \left[A_\bhat\,(D_\ghat)^4 T^{(3)}\right],
\label{scheme}
\ee
where the index contractions need to be worked out. Depending on the number of $D_\ahat$'s that act on $A_\bhat$, there can be several possible contributions, as we show in the following.

In our search for the possible contributions to (\ref{amplitude}), we are guided by dimensional analysis, 
SO($3,1$)$\times$SU(4)-invariance, the SYM equations of motion and gauge invariance. One can see  $\mathcal{M}_1$ is gauge-invariant because, under a gauge transformation $\delta A_\ahat = D_\ahat \Lambda$, one has
\be
\delta\mathcal{M}_1 \propto \Big\langle V_{(1)}\, \sqrt{z}\lambda^\ahat D_\ahat \Lambda\Big\rangle = -\,\Big\langle\Big(\sqrt{z}\lambda^\ahat D_\ahat V_{(1)}\Big)\,\Lambda\Big\rangle = 0\,,
\ee
since BRST-exact terms decouple and since $Q_{\frac{1}{2}}V_{(1)} = 0$ implies $\sqrt{z}\lambda^\ahat D_\ahat V_{(1)} = 0$. We also take into account that $T^{(3)}$ is a G-analytic superfield, such that independent contributions only contain $(\bu D_\ahat)$ derivatives acting on it.

When the five $D_\ahat$'s in (\ref{scheme}) act on $A_\bhat$, we get a dimension-3 superfield. So this term could in principle give a contribution proportional to $\partial_a\partial_b W_{ij}$ or $\partial_a F_{bc}$. We do not consider $\partial_a\partial_b A_c$ because this term is not gauge-invariant. It is easy to convince oneself that one cannot construct a non-zero term out of $\partial_a F_{bc}$, and hence
the only possible contribution is
\be
\bu^6 \left[(D_\ahat)^5 A_\bhat\right] (D_\ghat)^4 T^{(3)} \sim (\bu\bu)^{\ell m} \partial_{\beta\bdot} \partial_{\alpha\adot} W_{\ell m}\, \bD^{\bdot n} \tbD{}^\adot_n \tD^{\alpha k} D^\beta_k T^{(3)}\,,
\ee
where we used the definition $\partial_{\alpha\adot} := (\sigma^a)_{\alpha\adot}\partial_a$.

When four $D_\ahat$'s hit $A_\bhat$, we get a dimension-$\frac{5}{2}$ superfield. This could be either $\partial_a W^{\beta j}$ or its conjugate, $\partial_a \bar{W}^{\bdot}_j$. Keeping in mind the SYM equation of motion $\partial_{\alpha\adot}W^{\alpha i}=0$, we are left with two possibilities:
\ba
\bu^6 \left[(D_\ahat)^4 A_\bhat\right] (D_\ghat)^5 T^{(3)}  & \sim &  \partial_{\alpha\adot} W^{\beta j}\, \tbD{}^\adot_k (\bD^k \tbD_j) \tD^{\alpha r} D_{\beta r} T^{(3)}\nonumber\\
  & & +\; \partial_{\alpha\adot} \bar{W}^\bdot_j \, \bD^r_\bdot \tbD{}^\adot_r (D_k \tD^j) \tD^{\alpha k} T^{(3)}\,.
\ea

Going on with this analysis, we find all the possible gauge-invariant
contributions coming from different numbers of derivatives acting on $A_\bhat$, until the last case:
\be
\bu^6 \left[D_\ahat A_\bhat\right] (D_\ghat)^8 T^{(3)} \sim W^{(1)}\, \bD^4 D^{\prime 4} T^{(3)}\,.
\ee
Note that, since $\mathcal{M}_1$ is gauge-invariant, any contributions coming partly from a term in which no $D_\ahat$ acts on $A_\bhat$ can be expressed as a linear combination of 
gauge-invariant terms in which at least one $D_\ahat$ acts on $A_\bhat$ .

In summary, we get the following list of possible terms:
\begin{enumerate}

\item $W^{(1)}\, \bD^4 D^{\prime 4} T^{(3)}\,,$ 

\item $(uu)_{\ell m} W_\beta^\ell \,\bD^4 D_k^\beta (\tD^k \tD^m) T^{(3)}\,,$

\item $(uu)^{\ell m}\bar{W}^\bdot_\ell\, (\tbD_k \tbD_m) \bD_\bdot^k D^{\prime 4} T^{(3)}\,,$

\item $(\sigma^{ab})_\gamma{}^\beta F_{ab}\, \bD^4 D^\gamma_j \tD^j_\beta T^{(3)} =: F_\gamma{}^\beta\, \bD^4 D^\gamma_j \tD^j_\beta T^{(3)}\,,$

\item $(\tilde{\sigma}^{ab})^\bdot{}_\adot F_{ab}\, \tbD_{\bdot j} \bD^{\adot j} D^{\prime 4} T^{(3)} =: F^\bdot{}_\adot\, \tbD_{\bdot j} \bD^{\adot j} D^{\prime 4} T^{(3)}\,,$

\item $(uu)^{mj}\partial_{\alpha\adot} W_{\ell m}\, (\bD^n\tbD_j) \tbD{}^\adot_n \tD^{\alpha k} (D_k \tD^\ell) T^{(3)}\,,$

\item $\partial_{\alpha\adot} W^{\beta j}\, \tbD{}^\adot_k (\bD^k \tbD_j) \tD^{\alpha r} D_{\beta r} T^{(3)}\,,$

\item $\partial_{\alpha\adot} \bar{W}^\bdot_j \, \bD^r_\bdot \tbD{}^\adot_r (D_k \tD^j) \tD^{\alpha k} T^{(3)}\,,$

\item $(\bu\bu)^{\ell m} \partial_{\beta\bdot} \partial_{\alpha\adot} W_{\ell m}\, \bD^{\bdot n} \tbD{}^\adot_n \tD^{\alpha k} D^\beta_k T^{(3)}\,.$

\end{enumerate}
%

To conclude this subsubsection, let us argue  that
$$
\bar{D}^4 D^{\prime 4} \,\big[W^{(1)}(x,\theta,u)\, T^{(3)}(x,\theta,u,\bu)\big]
$$
is also given by a linear combination of the terms listed above. This expression contains eight derivatives of the type $(\bu D_\ahat)$. When all these derivatives hit $T^{(3)}$, one obviously gets the first term in the list. If only one derivative acts on $W^{(1)}$, the resulting expression is proportional to either the second or the third term in the list, depending on the chirality of the derivative ($D_{\alpha i}$ or $\bD_\adot^i$). And so on until the case in which four $(\bu D_\ahat)$'s hit $W^{(1)}$  to give either zero or something proportional to the last term. Acting with five or more derivatives on $W^{(1)}$ gives zero, as can be seen from the fact that $(\bu D_\ahat)^5 W^{(1)}$  has U(1) charge $-\frac{3}{2}$, while the SYM fields $\{\phi_{ij}, \xi^{\alpha i}, \bar{\xi}^\adot_i, f_{ab}\}$ have U(1) charges ranging from $-1$ to $+1$.


\subsubsection{Proof using supersymmetry}

Defining $\mathbf{t}_n$ ($n = 1, \ldots, 9$) to be the $n$-th possible term as listed at the end of the previous subsubsection, the amplitude we are computing must have the form
\be
\mathcal{M}_1 = \int \d^4 x \int\d u \sum_{n=1}^9 c_n\,\mathbf{t}_n\,,
\label{termm}
\ee
for some constants $c_1, \ldots, c_9$. In this subsection, we will prove that these constants are uniquely determined (up to an overall factor) by supersymmetry. Because both the left and right-hand sides of (\ref{main}) are supersymmetric, this proves our conjecture since
both the left and right-hand sides of (\ref{main}) must be proportional to (\ref{termm}). Note that the left-hand side of (\ref{main}) is supersymmetric (up to total derivatives), because $\lambda^\ahat A_\ahat\, \widetilde{\Omega}_{(0)}T^{(3)}$ is annihilated by $\lambda^\ahat D_\ahat$ so BRST invariance of the pure spinor measure factor implies supersymmetry as usual \cite{nathan}. To see that the right-hand side of (\ref{main}) is also supersymmetric (up to total derivatives), it suffices to write the supersymmetry generators as $q_\ahat \sim D_\ahat + \partial/\partial x$ and note that $D_{\alpha i}  \bD^4 D^{\prime 4}  {\mathcal{F}}$ and $\bD_\adot^j  \bD^4 D^{\prime 4} {\mathcal{F}}$ vanish for any G-analytic superfield ${\mathcal{F}}$ (since there are only four independent $(\bu D)_{\alpha \Ap}$ and they are fermionic).

In order to uniquely determine the  constants $c_1, \ldots, c_9$, we have to impose supersymmetry which implies that $D_\ahat  \left(\sum_{n=1}^9 c_n\,\mathbf{t}_n\right) = 0$. We begin by acting on the possible terms with $D_{\alpha i}$. We have:

\begin{enumerate}

\item \begin{eqnarray*}
D_{\alpha i}\left[W^{(1)}\, \bD^4 D^{\prime 4} T^{(3)}\right]  & = &  \left[D_{\alpha i} W^{(1)}\right]\bD^4 D^{\prime 4} T^{(3)} + W^{(1)}\, D_{\alpha i} \bD^4 D^{\prime 4} T^{(3)}    \nonumber\\
 & = &   -2\,(uu)_{ij} W_\alpha^j \bD^4 D^{\prime 4} T^{(3)} + 8\, \partial_{\alpha\adot} W^{(1)}\,(\tbD_i \tbD_j) \bD^{\adot j} D^{\prime 4} T^{(3)}\,,
\end{eqnarray*}
where we used $D_{\alpha i} W_{jk} = -\varepsilon_{ijk\ell} W_\alpha^\ell$ and $[D_{\alpha i}, \bD^4] = -8\, (\tbD_i \tbD_j) \bD^{\adot j} \partial_{\alpha\adot}$ and integrated by parts.

\item \begin{eqnarray*}
D_{\alpha i} \left[(uu)_{\ell m} W_\beta^\ell \,\bD^4 D_k^\beta (\tD^k \tD^m) T^{(3)} \right]   & = &   \left[D_{\alpha i} W_\beta^\ell\right] \bD^4 D_k^\beta (\tD^k D_\ell) T^{(3)} - W_\beta^\ell\, D_{\alpha i}\bD^4 D_k^\beta (\tD^k D_\ell) T^{(3)}\\
   & = &   \frac{1}{2} F_\alpha{}^\beta\, \bD^4 D_{\beta k} (\tD^k D_i) T^{(3)} +\frac{1}{4} (uu)_{ij} W_\alpha^j \bD^4 D^{\prime 4} T^{(3)}\\
  & &   -\; 8\, \partial_{\alpha\adot} W_\beta^\ell\, (\tbD_i \tbD_j) \bD^{\adot j} D_k^\beta (\tD^k D_\ell) T^{(3)}\,,
\end{eqnarray*}
where we used $D_{\alpha i} W^{\beta \ell} =  \frac{1}{2} \delta_i^\ell F_\alpha{}^\beta$ and $$D_{\alpha i} D_k^\beta (\tD^k D_\ell) T^{(3)} = -\frac{1}{4}\delta_\alpha^\beta (uu)_{i\ell} D^{\prime 4} T^{(3)}\,.$$

\item \begin{eqnarray*}
D_{\alpha i} \left[ (uu)^{\ell m}\bar{W}^\bdot_\ell\, (\tbD_k \tbD_m) \bD_\bdot^k D^{\prime 4} T^{(3)}\right]   & = &   
(uu)^{\ell m}\left[ D_{\alpha i} \bar{W}^\bdot_\ell\right] (\tbD_k \tbD_m) \bD_\bdot^k D^{\prime 4} T^{(3)}\\
  & &   -\; (uu)^{\ell m}\bar{W}^\bdot_\ell\, D_{\alpha i} (\tbD_k \tbD_m) \bD_\bdot^k D^{\prime 4} T^{(3)}\\
  & = &   -2\, (uu)^{\ell m}\partial_{\alpha\adot} W_{i\ell}\, (\tbD_k \tbD_m) \bD^{\adot k} D^{\prime 4} T^{(3)}\\
  & &   -\; 2\, (uu)^{\ell m}\partial_{\alpha\adot} \bar{W}_\ell^\bdot \, \tbD{}^\adot_m \tbD_{\bdot i} D^{\prime 4} T^{(3)}\\
  & &    +\; 2\,(uu)^{\ell m}(\bu\bu)_{mi}\,\partial_{\alpha\adot} \bar{W}^\bdot_\ell\, \tbD{}^\adot_k \bD^k_\bdot D^{\prime 4}T^{(3)}\\
  & = &   -2\, (uu)^{\ell m}\partial_{\alpha\adot} W_{i\ell}\, (\tbD_k \tbD_m) \bD^{\adot k} D^{\prime 4} T^{(3)}\\
  & &   +\;3\,(uu)^{\ell m}(\bu\bu)_{mi}\,\partial_{\alpha\adot} \bar{W}^\bdot_\ell\, \tbD{}^\adot_k \bD^k_\bdot D^{\prime 4}T^{(3)}\,,
\end{eqnarray*}
where we used $D_{\alpha i} \bar{W}_{\adot j} = -2\, \partial_{\alpha\adot} W_{ij}$ and the identities $(\bu\bu)^{i[j}(\bu\bu)^{k\ell]} = 0$ and $\varepsilon^{\adot[\bdot}\varepsilon^{\gdot\dotd]} = 0$.

\item \begin{eqnarray*}
D_{\alpha i} \left[F_\gamma{}^\beta\, \bD^4 D^\gamma_j \tD^j_\beta T^{(3)} \right]   & = &   F_\gamma{}^\beta\, \bD^4 D_{\alpha i}D^\gamma_j \tD^j_\beta T^{(3)} + 8\, \partial_{\alpha\adot} F_\gamma{}^\beta\,(\tbD_i \tbD_j) \bD^{\adot j} D_k^\gamma \tD_\beta^k T^{(3)}\,,
\end{eqnarray*}
where we used $D_{\alpha i} F_\gamma{}^\beta = 0$.

\item \begin{eqnarray*}
D_{\alpha i} \left[F^\bdot{}_\adot\, \tbD_{\bdot j} \bD^{\adot j} D^{\prime 4} T^{(3)}\right]  & = &    -4\,\partial_{\alpha\adot} \bar{W}_i^\bdot\, \tbD_{\bdot j} \bD^{\adot j} D^{\prime 4} T^{(3)}\,,
\end{eqnarray*}
where we used $D_{\alpha i} F^\bdot{}_\adot = -4\,\partial_{\alpha\adot}\bar{W}^\bdot_i$ and $\partial_{\alpha\bdot} F^\bdot{}_\adot = 0$.

\item 
$$ D_{\alpha i} \left[(uu)^{mj}\partial_{\beta\bdot} W_{\ell m}\, (\bD^n\tbD_j) \tbD{}^\bdot_n \tD^{\beta k} (D_k \tD^\ell) T^{(3)} \right]  = \qquad\qquad\qquad\qquad\qquad
$$
\begin{eqnarray*}
  &=&   -\varepsilon_{i\ell mp}(uu)^{mj}\partial_{\beta\bdot}W_\alpha^p\,(\bD^n\tbD_j) \tbD{}^\bdot_n \tD^{\beta k} (D_k \tD^\ell) T^{(3)}\\
  &  &    +\;(uu)^{mj}\partial_{\beta\bdot} W_{\ell m}\, D_{\alpha i}(\bD^n\tbD_j) \tbD{}^\bdot_n \tD^{\beta k} (D_k \tD^\ell) T^{(3)}\\
  &=&   -\varepsilon_{i\ell mp}(uu)^{mj}\partial_{\beta\bdot}W_\alpha^p\,(\bD^n\tbD_j) \tbD{}^\bdot_n \tD^{\beta k} (D_k \tD^\ell) T^{(3)}\\
  &  &     -\;3\, (uu)^{mj}(\bu\bu)_{ji}\partial_{\alpha\adot} \partial_{\beta\bdot} W_{\ell m}\, \bD^{\adot n} \tbD{}^\bdot_n \tD^{\beta k} (D_k\tD^\ell) T^{(3)}\\
   &  &     +\;\frac{1}{4} (uu)^{mj}(\bu\bu)^{\ell p}(uu)_{pi} \partial_{\alpha\adot}W_{\ell m}\, (\bD^n \tbD_j)\tbD{}^\adot_n D^{\prime 4} T^{(3)}\,.
\end{eqnarray*}

\item \begin{eqnarray*}
D_{\alpha i}\left[\partial_{\beta\bdot} W^{\gamma j}\, \tbD{}^\bdot_k (\bD^k \tbD_j) \tD^{\beta r} D_{\gamma r} T^{(3)}\right]   & = &    \left[D_{\alpha i}\partial_{\beta\bdot} W^{\gamma j}\right]\tbD{}^\bdot_k (\bD^k \tbD_j) \tD^{\beta r} D_{\gamma r} T^{(3)} \\
  &  &     -\; \partial_{\beta\bdot} W^{\gamma j}\, D_{\alpha i}\tbD{}^\bdot_k (\bD^k \tbD_j) \tD^{\beta r} D_{\gamma r} T^{(3)}\\
  & = &     \frac{1}{2} \partial_{\beta\bdot} F_\alpha{}^\gamma\, \tbD{}^\bdot_k (\bD^k \tbD_i) \tD^{\beta r} D_{\gamma r} T^{(3)}\\
  &  &    +\; 3\, (\bu\bu)_{ij}\partial_{\alpha\adot} \partial_{\beta\bdot}W^{\gamma j}\, \tbD{}^\bdot_k \bD^{\adot k}\tD^{\beta r} D_{\gamma r} T^{(3)}\\
  &  &     +\; \partial_{\beta\bdot}W^{\gamma j}\, \tbD{}^\bdot_k (\bD^k \tbD_j) D_{\alpha i}\tD^{\beta r} D_{\gamma r} T^{(3)}\,.
\end{eqnarray*}

\item \begin{eqnarray*}
D_{\alpha i} \left[\partial_{\beta\bdot} \bar{W}^\gdot_j \, \bD^r_\gdot \tbD{}^\bdot_r (D_k \tD^j) \tD^{\beta k} T^{(3)}\right]   & = &    -2\, \partial_{\alpha\adot} \partial_{\beta\bdot} W_{ij}\, \bD^{\adot r}\tbD{}^\bdot_r (D_k \tD^j) \tD^{\beta k} T^{(3)}\\
  &  &    +\;\frac{1}{4} (\bu\bu)^{jm}(uu)_{mi} \partial_{\alpha\adot}\bar{W}^\gdot_j \, \bD^r_\gdot \tbD{}^\adot_r D^{\prime 4} T^{(3)}\,.
\end{eqnarray*}

\item \begin{eqnarray*}
D_{\alpha i} \left[ (\bu\bu)^{\ell m} \partial_{\beta\bdot} \partial_{\gamma\gdot} W_{\ell m}\, \bD^{\bdot n} \tbD{}^\gdot_n \tD^{\gamma k} D^\beta_k T^{(3)}\right]   & = &   -2\, (\bu\bu)_{ij} \partial_{\beta\bdot} \partial_{\gamma\gdot} W_\alpha^j\,\bD^{\bdot n} \tbD{}^\gdot_n \tD^{\gamma k} D^\beta_k T^{(3)}\\
  &  &   +\; (\bu\bu)^{\ell m} \partial_{\beta\bdot} \partial_{\gamma\gdot} W_{\ell m}\, \bD^{\bdot n} \tbD{}^\gdot_n \tD^{\gamma k} D^\beta_k D_{\alpha i}T^{(3)}\,.
\end{eqnarray*}

\end{enumerate}

We see that acting with $D_{\alpha i}$ on $\mathbf{t}_n$ produces various terms. In order to impose that the amplitude is supersymmetric, we need to collect the terms which should cancel independently. In the following, we organize the results according to the superfields appearing in each term. The imposition that they cancel gives rise to a system of many equations for the constants $c_n$, which have to be solved at the same time.

\begin{itemize}

\item Terms proportional to $W^{\beta j}$ (without $x$-derivatives):
\begin{eqnarray*}
{\left. \left(D_{\alpha i} {\mathcal{M}}_1\right)\right|}_{W^{\beta j}}   & = &    \left(\frac{1}{4}c_2 -2c_1\right)(uu)_{ij} W_\alpha^j \bD^4 D^{\prime 4} T^{(3)}\,.
\end{eqnarray*}
Hence we get our first equation:
\be
\frac{1}{4}c_2 -2c_1 = 0\,.
\label{firsteq}
\ee

\item Terms proportional to $F_\gamma{}^\beta$:
\begin{eqnarray*}
{\left. \left(D_{\alpha i} {\mathcal{M}}_1\right)\right|}_{F_\gamma{}^\beta}   & = &   \frac{1}{2}c_2\,F_\alpha{}^\beta\, \bD^4 D_{\beta k} (\tD^k D_i) T^{(3)} + c_4\, F_\gamma{}^\beta\, \bD^4 D_{\alpha i}D^\gamma_j \tD^j_\beta T^{(3)}\\
  & = &    \left(\frac{1}{2}c_2 + \frac{2}{3}c_4\right)F_\alpha{}^\beta\, \bD^4 D_{\beta k} (\tD^k D_i) T^{(3)}\,,
\end{eqnarray*}
where we used $F_\gamma{}^\beta \bD^4 D_{\alpha i} D^\gamma_j \tD^j_\beta T^{(3)} = \frac{2}{3}F_\alpha{}^\beta \bD^4 D_{\beta k} (\tD^k D_i) T^{(3)}$. Therefore,
\be
\frac{1}{2}c_2 + \frac{2}{3}c_4 = 0\,.
\ee

\item Terms proportional to $\partial F_\gamma{}^\beta$:
\begin{eqnarray*}
{\left. \left(D_{\alpha i} {\mathcal{M}}_1\right)\right|}_{\partial F_\gamma{}^\beta}   & = &   8c_4\, \partial_{\alpha\adot} F_\gamma{}^\beta\,(\tbD_i \tbD_j) \bD^{\adot j} D_k^\gamma \tD_\beta^k T^{(3)}\\
  &  &    +\;\frac{1}{2}c_7\,\partial_{\beta\bdot} F_\alpha{}^\gamma\, \tbD{}^\bdot_k (\bD^k \tbD_i) \tD^{\beta r} D_{\gamma r} T^{(3)}\\
  & = &    \left(8c_4 + \frac{1}{2}c_7\right)\partial_{\alpha\adot} F_\gamma{}^\beta\,(\tbD_i \tbD_j) \bD^{\adot j} D_k^\gamma \tD_\beta^k T^{(3)}\,,
\end{eqnarray*}
where we used $\partial_{\alpha\adot} F_\beta{}^\gamma = \partial_{\beta\adot} F_\alpha{}^\gamma$. So we get
\be
8c_4 + \frac{1}{2}c_7 = 0\,.
\ee

\item Terms proportional to $\partial \bar{W}^\adot_j$:
\begin{eqnarray*}
{\left. \left(D_{\alpha i} {\mathcal{M}}_1\right)\right|}_{\partial \bar{W}^\adot_j}   & = &   3c_3\,(uu)^{\ell m}(\bu\bu)_{mi}\,\partial_{\alpha\adot} \bar{W}^\bdot_\ell\, \tbD{}^\adot_k \bD^k_\bdot D^{\prime 4}T^{(3)}\\
  & &    -\;4c_5\,\partial_{\alpha\adot} \bar{W}_i^\bdot\, \tbD_{\bdot j} \bD^{\adot j} D^{\prime 4} T^{(3)}\\
  & &   +\;\frac{1}{4}c_8\,(\bu\bu)^{jm}(uu)_{mi} \partial_{\alpha\adot}\bar{W}^\gdot_j \, \bD^r_\gdot \tbD{}^\adot_r D^{\prime 4} T^{(3)}\\
  & = &   \left(3c_3 - 4c_5\right)(uu)^{\ell m}(\bu\bu)_{mi}\,\partial_{\alpha\adot} \bar{W}^\bdot_\ell\, \tbD{}^\adot_k \bD^k_\bdot D^{\prime 4}T^{(3)}\\
  & &    - \left(\frac{1}{4}c_8 +4c_5\right)(\bu\bu)^{\ell m}(uu)_{mi}\,\partial_{\alpha\adot} \bar{W}^\bdot_\ell\, \tbD{}^\adot_k \bD^k_\bdot D^{\prime 4}T^{(3)}\,,
\end{eqnarray*}
where we used $(uu)^{\ell m}(\bu\bu)_{mi} + (\bu\bu)^{\ell m}(uu)_{mi} = \delta_i^\ell$. Thus we obtain two equations:
\be
\left\{\begin{array}{rcl}
3c_3 - 4c_5 & = & 0\,, \\
\displaystyle\frac{1}{4}c_8 +4c_5 & = & 0\,.
\end{array}
\right.
\ee

\item Terms proportional to $\partial W^{\beta j}$:
\begin{eqnarray*}
{\left. \left(D_{\alpha i} {\mathcal{M}}_1\right)\right|}_{\partial W^{\beta j}}   & = &   -8c_2\, \partial_{\alpha\adot} W_\beta^\ell\, (\tbD_i \tbD_j) \bD^{\adot j} D_k^\beta (\tD^k D_\ell) T^{(3)}\\
  &  &    -\;c_6\,\varepsilon_{i\ell mp}(uu)^{mj}\partial_{\beta\bdot}W_\alpha^p\,(\bD^n\tbD_j) \tbD{}^\bdot_n \tD^{\beta k} (D_k \tD^\ell) T^{(3)}\\
  &  &     +\;c_7\,\partial_{\beta\bdot}W^{\gamma j}\, \tbD{}^\bdot_k (\bD^k \tbD_j) D_{\alpha i}\tD^{\beta r} D_{\gamma r} T^{(3)}\\
  & = &   \left(c_6 - 8c_2\right) \partial_{\alpha\adot} W_\beta^\ell\, (\tbD_i \tbD_j) \bD^{\adot j} D_k^\beta (\tD^k D_\ell) T^{(3)}\\
  &  &    + \left(\frac{3}{2}c_6 - c_7\right) \partial_{\gamma\adot} W_\beta^\ell\, (\tbD_\ell \tbD_j)\bD^{\adot j} \tD^{\gamma k} D_k^\beta D_{\alpha i} T^{(3)}\,,
\end{eqnarray*}
where we used the identities $\varepsilon_{i\ell mp} = -3\, (uu)_{[i\ell}(\bu\bu)_{m]p} - 3\, (\bu\bu)_{[i\ell}(uu)_{m]p}$, $\varepsilon_{\alpha[\beta}\varepsilon_{\gamma\delta]}=0$ and $(uu)_{[i\ell}(uu)_{m]p}=0$. Hence we get two more equations:
\be
\left\{\begin{array}{rcl}
c_6 - 8c_2 & = & 0\,, \\
\displaystyle\frac{3}{2}c_6 - c_7 & = & 0\,.
\end{array}
\right.
\ee

\item Terms proportional to $\partial^2 W^{\beta j}$:
\begin{eqnarray*}
{\left. \left(D_{\alpha i} {\mathcal{M}}_1\right)\right|}_{\partial^2 W^{\beta j}}   & = &   3c_7\,(\bu\bu)_{ij}\partial_{\alpha\adot} \partial_{\beta\bdot}W^{\gamma j}\, \tbD{}^\bdot_k \bD^{\adot k}\tD^{\beta r} D_{\gamma r} T^{(3)}\\
   &  &     -\;2c_9\,(\bu\bu)_{ij} \partial_{\beta\bdot} \partial_{\gamma\gdot} W_\alpha^j\,\bD^{\bdot n} \tbD{}^\gdot_n \tD^{\gamma k} D^\beta_k T^{(3)}\\
  & = &   \left(3c_7 + 2c_9\right) (\bu\bu)_{ij}\partial_{\alpha\adot} \partial_{\beta\bdot}W^{\gamma j}\, \tbD{}^\bdot_k \bD^{\adot k}\tD^{\beta r} D_{\gamma r} T^{(3)}\,,
\end{eqnarray*}
whence
\be
3c_7 + 2c_9 = 0\,.
\ee

\item Terms proportional to $\partial W_{ij}$:
\begin{eqnarray*}
{\left. \left(D_{\alpha i} {\mathcal{M}}_1\right)\right|}_{\partial W{ij}}   & = &   8c_1\,\partial_{\alpha\adot} W^{(1)}\,(\tbD_i \tbD_j) \bD^{\adot j} D^{\prime 4} T^{(3)}\\
   &  &     -\;2c_3\,(uu)^{\ell m}\partial_{\alpha\adot} W_{i\ell}\, (\tbD_k \tbD_m) \bD^{\adot k} D^{\prime 4} T^{(3)}\\
   &  &     +\;\frac{1}{4}c_6\, (uu)^{mj}(\bu\bu)^{\ell p}(uu)_{pi} \partial_{\alpha\adot}W_{\ell m}\, (\bD^n \tbD_j)\tbD{}^\adot_n D^{\prime 4} T^{(3)}\\
  & = &   \left(8c_1 + c_3\right)\partial_{\alpha\adot} W^{(1)}\,(\tbD_i \tbD_j) \bD^{\adot j} D^{\prime 4} T^{(3)}\\
   &  &     + \left(2c_3 + \frac{1}{4}c_6\right) (uu)^{mj}(\bu\bu)^{\ell p}(uu)_{pi} \partial_{\alpha\adot}W_{\ell m}\, (\bD^n \tbD_j)\tbD{}^\adot_n D^{\prime 4} T^{(3)}\,,
\end{eqnarray*}
so that we must have
\be
\left\{\begin{array}{rcl}
8c_1 + c_3 & = & 0\,, \\
2c_3 + \displaystyle\frac{1}{4}c_6 & = & 0\,.
\end{array}
\right.
\ee

\item Terms proportional to $\partial^2 W_{ij}$:
\begin{eqnarray*}
{\left. \left(D_{\alpha i} {\mathcal{M}}_1\right)\right|}_{\partial^2 W{ij}}   & = &   -3c_6\,(uu)^{mj}(\bu\bu)_{ji}\partial_{\alpha\adot} \partial_{\beta\bdot} W_{\ell m}\, \bD^{\adot n} \tbD{}^\bdot_n \tD^{\beta k} (D_k\tD^\ell) T^{(3)}\\
    &  &    -\;2c_8\,\partial_{\alpha\adot} \partial_{\beta\bdot} W_{ij}\, \bD^{\adot r}\tbD{}^\bdot_r (D_k \tD^j) \tD^{\beta k} T^{(3)}\\
   &  &    +\;c_9\,(\bu\bu)^{\ell m} \partial_{\beta\bdot} \partial_{\gamma\gdot} W_{\ell m}\, \bD^{\bdot n} \tbD{}^\gdot_n \tD^{\gamma k} D^\beta_k D_{\alpha i}T^{(3)}\\
  & = &   \left(3c_6 - 2c_8\right) (uu)^{mj}(\bu\bu)_{ji}\partial_{\alpha\adot} \partial_{\beta\bdot} W_{m\ell}\, \bD^{\adot n} \tbD{}^\bdot_n \tD^{\beta k} (D_k\tD^\ell) T^{(3)}\\
   &  &    - \left(\frac{4}{3}c_9 + 2 c_8\right)(\bu\bu)^{mj}(uu)_{ji}\partial_{\alpha\adot} \partial_{\beta\bdot} W_{m\ell}\, \bD^{\adot n} \tbD{}^\bdot_n \tD^{\beta k} (D_k\tD^\ell) T^{(3)}\,,
\end{eqnarray*}
then we find the last pair of equations:
\be
\left\{\begin{array}{rcl}
3c_6 - 2c_8 & = & 0\,, \\
\displaystyle\frac{4}{3}c_9 +2 c_8 & = & 0\,.
\end{array}
\right.
\label{lasteq}
\ee

\end{itemize}

Putting together (\ref{firsteq}), $\ldots$ , (\ref{lasteq}), we get the following system of equations:
\be
\left\{
\begin{array}{rcl}
\frac{1}{4}c_2 -2c_1 &=& 0 \\
\frac{1}{2}c_2 + \frac{2}{3}c_4 &=& 0 \\
8c_4 + \frac{1}{2}c_7 &=& 0 \\
3c_3 - 4c_5 & = & 0 \\
\frac{1}{4}c_8 +4c_5 & = & 0 \\
c_6 - 8c_2 & = & 0\ \\
\frac{3}{2}c_6 - c_7 & = & 0 \\
3c_7 + 2c_9 &=& 0 \\
8c_1 + c_3 & = & 0 \\
2c_3 + \frac{1}{4}c_6 & = & 0 \\
3c_6 - 2c_8 & = & 0 \\
\frac{4}{3}c_9 + 2c_8 & = & 0\,.
\end{array}
\right.
\ee
We see there are a few more equations than unknowns, but they turn out to be not all independent. 
Setting $c_1 = 1$ as our normalization,\footnote{Here we are implicitly neglecting the case in which all the constants $c_n$ vanish, which of course is also a solution to the system of equations. It is straightforward to show the right-hand side of (\ref{main}) is non-vanishing, and the left-hand side can also be shown to be non-zero by direct computation of one of the possible terms, for example by choosing the gauge $A_\ahat = (\theta\gamma^{[ij]})_\ahat W_{ij}$ with $W_{ij}$ constant.}  this system can be  solved to give
\ba
c_2 = -c_3   & = &   8\,,\nonumber\\
c_4 = c_5   & = &  -6\,,\nonumber\\
c_6   & = &   64\,,\\
c_7 = c_8   & = &   96\,,\nonumber\\
c_9   & = &   -144\,.\nonumber
\ea
Note that ${\mathcal{M}}_1$ is real, which implies $D_{\alpha i}{\mathcal{M}}_1 = 0 \; \Longleftrightarrow \; \bD_\adot^i {\mathcal{M}}_1 = 0$.

Hence we have proved that, up to an overall factor, there is only one combination of the possible terms which is supersymmetric. This in turn shows that equation (\ref{main}) is indeed true.

\subsection{\boldmath{The cases $N > 1$}}

In the last subsection, we have shown that
$$
V_{(1)} = z \int \d u\,\Omega_{(0)} T^{(3)}(x,\theta,u,\bu)
$$
implies
\be
{\mathcal{M}}_1 = {\Big\langle V_\mathrm{SYM}\,V_{(1)}\Big\rangle}_{\mathrm{D3-brane}} \propto \int\d^4x\int\d u\,\bD^4 D^{\prime 4}\left[W^{(1)}\, T^{(3)}\right].
\label{m1}
\ee
Now we want to generalize this result to any $N$, i.e. we want to show
\ba
 {\mathcal{M}}_N  & := &  \int \d\xi_1\ldots\d\xi_{N-1}\,{\Big\langle V_\mathrm{SYM}\,V_{(N)}\,U_\mathrm{SYM}(\xi_1)\ldots U_\mathrm{SYM}(\xi_{N-1})\Big\rangle}_{\mathrm{D3-brane}}  \nonumber\\
 & \propto &   \int\d^4x\int\d u\,\bD^4 D^{\prime 4}\left[W^{(N)}\, T^{(4-N)}\right], \label{mn}
\ea
where we recall $U_\mathrm{SYM}$ is the integrated version of $V_\mathrm{SYM} = \sqrt{z}\,\lambda^\ahat A_\ahat$ and $V_{(N)}$ was defined in (\ref{vertex}).

\subsubsection{\boldmath{The $N = 2$ case}}

As a natural first step, let us analyze the case $N=2$. In this case, we want to compute the following scattering amplitude:
\be
{\mathcal{M}}_2 = \int\d\xi \,{\Big\langle V_\mathrm{SYM}(\infty)\,V_{(2)}(\mathrm{i}\epsilon,-\mathrm{i}\epsilon)\,U_\mathrm{SYM}(\xi)\Big\rangle}_{\mathrm{D3-brane}}\,,
\label{m2}
\ee
where we have fixed the worldsheet positions of the unintegrated vertex operators (here $\epsilon$ is a positive infinitesimal).

The easiest way to do this computation is by taking the OPE of $V_{(2)}(\mathrm{i}\epsilon,-\mathrm{i}\epsilon)$ with $U_\mathrm{SYM}(\xi)$ and looking for terms which can contribute to 
the dual to $V_\mathrm{SYM}$. Here we use the word ``dual'' meaning an object $O$ such that $\langle V_\mathrm{SYM}\,O\rangle$ is nonzero. This object must be in the ghost-number +2 cohomology of the BRST operator $Q$ and its product with $V_\mathrm{SYM}$ should include terms proportional to the measure factor
$$
 (\lambda\gamma^\mu\theta)(\lambda\gamma^\nu\theta)(\lambda\gamma^\rho\theta)(\theta\gamma_{\mu\nu\rho}\theta)\,.
$$

Moreover, the amplitude must of course be PSU($2,2|4$)-invariant, and in particular SU(4)-invariant.
A superfield in the $[0,p,0]$ representation of SU(4) should couple to another superfield in the same representation, so that a scalar ($[0,0,0]$) is present in their tensor product decomposition.
The Dynkin label $p$ is related to the number of $y$'s ($n_y$) in a vertex operator by $p = n_y + 1$, as can be seen from the coupling (\ref{couple}). The argument goes as follows: $T^{(4-N)}$ couples to $(W_{ij})^N$ and hence $V_{(N)}$ corresponds to $(W_{ij})^N$. Now, $(W_{ij})^N$ transforms in the $[0,N,0]$ of SU(4), while $V_{(N)} \sim y^{N-1}$. Therefore, since $V_\mathrm{SYM}$ is independent of $y$, so must  $O$ be.

As seen in subsection \ref{BRSTcohomology}, an object in the ghost-number +2 cohomology of $Q$ has the form
$$
O_{(N)} = z^{2-N} \int\d u \sum_{n=0}^4 8^n P_n(N)\,(yuu)^{N-n-1}\,\Omega_{(n)}G^{(4-N)}(x,\theta,u,\bu)\,,
$$
where $G^{(4-N)}(x,\theta,u,\bu)$ is some G-analytic superfield  of harmonic U(1) charge $4-N$ and the operators $\Omega_{(n)}$ were defined in (\ref{omega}). Since, by the argument given in the previous paragraph, the dual to $V_\mathrm{SYM}$ must be independent of $y$, it follows that it can be written as
\be
O \equiv O_{(1)} = z\int\d u\,\Omega_{(0)}G^{(3)} + Q\chi\,,
\label{dual}
\ee
where we have explicitly included a possible BRST-trivial term. Thus our problem is equivalent to finding the superfield $G^{(3)}$.

To find terms in the OPE of $V_{(2)}(\mathrm{i}\epsilon,-\mathrm{i}\epsilon)$ with $U_\mathrm{SYM}(\xi)$ which can contribute to $O$, 
one first considers the OPE's coming from the conformal-weight $+1$ operators of the integrated vertex operator.
Since we are only interested in the $z \to 0$ limit, we can consider the flat-space expression for the integrated vertex operator corresponding to states propagating in a D3-brane world-volume, i.e.
\be
U_\mathrm{SYM} = \partial\theta^\ahat A_\ahat + (\partial x^a + \theta\gamma^a\partial\theta)A_a + z\partial y^{ij} W_{ij}    + \frac{1}{2}zP_{\hat{\psi}^\ahat} W^\ahat + \mathcal{O}(z^2)\,,
\label{Usym}
\ee
where $P_{\hat{\psi}^\ahat}$ is the momentum conjugate to ${\hat{\psi}^\ahat}$. The term $\frac{1}{2}zP_{\hat{\psi}^\ahat} W^\ahat$ is just the $d_\ahat W^\ahat$ term written in AdS${}_5 \times \mathrm{S}^5$ notation. The easiest way to see $P_{\hat{\psi}^\ahat}$ corresponds to $d_\ahat$ in the $z\to0$ limit is by recalling the expression for $Q_{-\frac{1}{2}}$, the lowest term in the $Q$ expansion in powers of $z$, which is reproduced below:
$$
Q_{-\frac{1}{2}} \propto \frac{1}{\sqrt{z}}\left(\lambda^{+ \gamma m} y_{mi} P_{\psi_i^\gamma} + \blambda^{+\adot}_j y^{ji} P_{\bar{\psi}^{i\adot}}\right). 
$$
Since this expression can be written in ten-dimensional notation as $z^{-\frac{1}{2}}\,\lambda^{+\ahat}P_{\hat{\psi}^\ahat}$ and the expression in flat space would be $\lambda^\ahat d_\ahat$, it follows that $P_{\hat{\psi}^\ahat}$ corresponds to $d_\ahat$. The factor of $z$ in $\frac{1}{2}zP_{\hat{\psi}^\ahat} W^\ahat$ enters by dimensional analysis, and the numerical factor is needed for BRST-invariance.

The terms of order $z^2$ in (\ref{Usym}) will not contribute. For example, one of these terms is $z^2 N^{ab}F_{ab}$. Because the OPE of $N^{ab}$ and $\lambda$ is independent of $z$, this term cannot give a contribution of order $z$, and therefore cannot contribute to the dual of $V_{\mathrm{SYM}}$  (cf. (\ref{dual})) in the $z \to 0$ limit.

Another term  in $U_\mathrm{SYM}$ which will not contribute is $\partial x^a A_a$. Since the kinetic term for $x^a$ in the Lagrangian is
$$
\frac{1}{2}\frac{\partial x^a \,\bar{\partial}x_a}{z^2}\,,
$$
it turns out the OPE of $\partial x^a$ and a superfield depending on $x$ is also of order $z^2$.

In fact, since the dual to $V_\mathrm{SYM}$ does not depend on $y$ or $\hat{\psi}$, only the terms  $z\,\partial y^{ij}W_{ij}$ and $\frac{1}{2}zP_{\hat{\psi}^\ahat} W^\ahat$ in $U_\mathrm{SYM}$ may contribute to  ${\mathcal{M}}_2$. The reason is that these are the only terms which can remove the $y$- and $\hat{\psi}$-dependence of $V_{(2)}$ via the OPE's
\be
P_{\hat{\psi}^\ahat}(\xi)\,\hat{\psi}^\bhat(\zeta,\bar{\zeta}) \sim \frac{\delta_\ahat^\bhat}{\xi-\zeta} + \frac{\delta_\ahat^\bhat}{\xi-\bar{\zeta}}
\label{ope2}
\ee
and
\be
\partial y^{ij}(\xi)\,y^{k\ell}(\zeta, \bar{\zeta}) \sim \frac{2\varepsilon^{ijk\ell}}{\xi - \zeta} + \frac{2\varepsilon^{ijk\ell}}{\xi - \bar{\zeta}} +\cdots\,,
\label{ope}
\ee
where the dots include terms depending on $y$.

The closed superstring vertex operator for $N=2$ is given by (cf. (\ref{vertex}))
\be
V_{(2)} = \int \d u \left[(yuu)\,\Omega_{(0)} + 8\,\Omega_{(1)}\right]T^{(2)}\,.
\ee
Hence, there are two contributions to the amplitude. The $\Omega_{(0)}$-term in $V_{(2)}$ is contracted with the $z\,\partial y^{ij}W_{ij}$ in  $U_\mathrm{SYM}$ to give
 $$
4\left[\frac{1}{\xi-\mathrm{i}\epsilon}+\frac{1}{\xi+\mathrm{i}\epsilon}\right]z\int\d u\,W^{(1)}\,\Omega_{(0)}T^{(2)}\,,
$$
while the $\Omega_{(1)}$-term in $V_{(2)}$ is contracted with the $\frac{1}{2}zP_{\hat{\psi}^\ahat} W^\ahat$ in  $U_\mathrm{SYM}$ to give
$$
-4\left[\frac{1}{\xi-\mathrm{i}\epsilon}+\frac{1}{\xi+\mathrm{i}\epsilon}\right]z\int\d u\,\widehat{\Omega}_{(1)} T^{(2)}\,,
$$
where $\widehat{\Omega}_{(n)}$ is equal to $\Omega_{(n)}$ with the substitution $\hat{\psi}^\ahat \longmapsto W^\ahat$.

So, performing the integral over $\d\xi$ in the complex plane, choosing a contour that encloses the pole $\xi = \mathrm{i} \epsilon$, we obtain
\be
{\mathcal{M}}_2 \propto \Big\langle V_\mathrm{SYM}\, z \int\d u\left[W^{(1)}\,\Omega_{(0)} - \widehat{\Omega}_{(1)}\right] T^{(2)}\Big\rangle\,.
\label{m2omega1}
\ee

Since $\sqrt{z} \lambda^\ahat D_\ahat (\lambda\gamma^\mu W) = 0$, the $\widehat{\Omega}_{(n)}$'s also satisfy the equations (\ref{omegahat}), with the substitution $\hat{\psi}^\ahat \longmapsto W^\ahat$.
Hence, it is not difficult to show that
\be
z \int\d u\left[W^{(1)}\,\Omega_{(0)} - \widehat{\Omega}_{(1)}\right] T^{(2)}\,,
\label{extraterm}
\ee
and thus the amplitude,
is BRST-invariant as it should be. Let us do that. First, note that the $\left(2z\frac{\partial}{\partial z} + y^{k\ell}\frac{\partial}{\partial y^{k\ell}} - \lambda^\ahat\frac{\partial}{\partial \lambda^\ahat}             \right)$-part of $Q_{\frac{1}{2}}$ annihilates (\ref{extraterm}). Then, since this expression does not depend on $y$, it is left to show that (\ref{extraterm}) is annihilated by $\sqrt{z} \lambda^\ahat D_\ahat + \tilde{w}^\ahat r_\ahat$. Using (\ref{omegahat.a}) and the SYM equation of motion $D_\ahat W_{ij} = -(\gamma_{[ij]} W)_\ahat$, we get
\be
\left(\sqrt{z} \lambda^\ahat D_\ahat + \tilde{w}^\ahat r_\ahat\right) (W^{(1)}\,\Omega_{(0)}T^{(2)}) = -\sqrt{z}\,(\lambda\gamma_{[ij]}W)(uu)^{ij}\Omega_{(0)}T^{(2)}\,,
\ee
while the modified version of (\ref{omegahat.b}) gives
\be
\left(\sqrt{z} \lambda^\ahat D_\ahat + \tilde{w}^\ahat r_\ahat\right)\widehat{\Omega}_{(1)}T^{(2)} = -\sqrt{z}\, (\lambda\gamma^{[ij]}W)(uu)_{ij}\Omega_{(0)}T^{(2)}\,.
\ee
Thus the BRST-variations of the two terms in (\ref{extraterm}) cancel each other, implying that expression  is indeed BRST-invariant.

Previously we argued that any object in the ghost-number $+2$ cohomology of $Q$ which does not depend on $y$ can be expressed in the form (\ref{dual}) for some $G^{(3)}$ which is G-analytic. Therefore, (\ref{extraterm})
can be expressed in the form (\ref{dual}).
Moreover, when $W_{ij}$ is constant, it is easy to see that (\ref{extraterm}) can be expressed in this form with $G^{(3)} = W^{(1)}\,T^{(2)}$. So $G^{(3)} = W^{(1)}\,T^{(2)} + f $ where $f$ is a G-analytic term involving derivatives of $W_{ij}$. But $D W^{(1)}$ is not G-analytic and there are no G-analytic terms of the appropriate dimension that can be constructed out of derivatives of $W_{ij}$. So $G^{(3)}$ must be equal to $W^{(1)}\,T^{(2)}$ even when $W_{ij}$ is not constant, i.e. (\ref{extraterm}) must be equal to
\be
z \int \d u\,\Omega_{(0)}\left[W^{(1)}\,T^{(2)}\right] + Q\chi_2\,,
\ee
where the BRST-trivial term $Q\chi_2$ vanishes when $W_{ij}$ is constant. This implies
\be
{\mathcal{M}}_2 \propto \Big\langle V_\mathrm{SYM}\, z \int\d u\,\Omega_{(0)}\left[W^{(1)}\,T^{(2)}\right]\Big\rangle .
\ee
Note this is consistent with the gauge transformation (\ref{symm}), since $\delta T^{(2)} = u_A{}^i\frac{\partial}{\partial \bu_\Ap{}^i}\Xi^{(1)\,A}_\Ap$ and the analyticity of $W^{(1)}$ imply 
$$
\delta {\mathcal{M}}_2 \propto z \int\d u\,\Omega_{(0)}\left[u_A{}^i\frac{\partial}{\partial \bu_\Ap{}^i}\left(W^{(1)}\,\Xi^{(1)\,A}_\Ap\right)\right],
$$
which is BRST-trivial.

Finally, using (\ref{m1}), we obtain
\be
{\mathcal{M}}_2 \propto \int\d^4x\int\d u\,\bD^4 D^{\prime 4}\left[W^{(1)}\, W^{(1)}\, T^{(2)}\right] = \int\d^4x\int\d u\,\bD^4 D^{\prime 4}\left[W^{(2)}\, T^{(2)}\right],
\ee
thus proving (\ref{mn}) in the case $N=2$.

\subsubsection{\boldmath{Generalization to any $N>1$}}

We are now in position to prove (\ref{mn}) for any $N$. Let us copy it here for the sake of readability:
\ba
 {\mathcal{M}}_N  & := &  \int \d\xi_1\ldots\d\xi_{N-1}\,{\Big\langle V_\mathrm{SYM}(\infty)\,V_{(N)}(\mathrm{i}\epsilon,-\mathrm{i}\epsilon)\,U_\mathrm{SYM}(\xi_1)\ldots U_\mathrm{SYM}(\xi_{N-1})\Big\rangle}_{\mathrm{D3-brane}}  \nonumber\\
 & \propto &   \int\d^4x\int\d u\,\bD^4 D^{\prime 4}\left[W^{(N)}\, T^{(4-N)}\right]. \nonumber
\ea
Again, we are looking for the dual to $V_\mathrm{SYM}$ in the form (\ref{dual}), i.e. we are looking for the expression of the G-analytic superfield $G^{(3)}$ in the case of the amplitude  ${\mathcal{M}}_N$.

As argued in the previous subsubsection, only the terms  $z\,\partial y^{ij}W_{ij}$ and $\frac{1}{2}zP_{\hat{\psi}^\ahat} W^\ahat$ in the integrated vertex operators can remove the $y$- and $\hat{\psi}$-dependence from the supergravity vertex operator through their OPE's  and thus contribute to  ${\mathcal{M}}_N$. This also implies that there can be no contribution coming from contractions between two or more integrated vertex operators. Recalling that
$$
V_{(N)} = z^{2-N} \int\d u \sum_{n=0}^4 8^n P_n(N)\,(yuu)^{N-n-1}\,\Omega_{(n)}T^{(4-N)}(x,\theta,u,\bu)\,,
\label{vertex}
$$
the OPE's give, after performing the ($N-1$) integrations over the $\d\xi$'s,
\be
{\mathcal{M}}_N \propto \Big\langle V_\mathrm{SYM}\, z \int\d u \sum_{n=0}^4 (-1)^n P_n(N)\,W^{(N-n-1)}\,\widehat{\Omega}_{(n)}T^{(4-N)}(x,\theta,u,\bu)\Big\rangle\,.
\ee

Note the factor of $z$ comes from the product of each $z$ in the $N-1$ integrated vertex operators with the $z^{2-N}$ factor of $V_{(N)}$. Again, one can use equations (\ref{omegahat}) with $\Omega \longmapsto \widehat{\Omega}$ and  $\hat{\psi}^\ahat \longmapsto W^\ahat$ to show
\be
z \int\d u \sum_{n=0}^4 (-1)^n P_n(N)\,W^{(N-n-1)}\,\widehat{\Omega}_{(n)}T^{(4-N)}(x,\theta,u,\bu)
\label{fim}
\ee
is BRST-invariant. The calculation is similar to the one performed below (\ref{extraterm}), but with more terms.  Then, by arguments completely analogous to the ones given at the end of the previous subsubsection, we conclude that (\ref{fim}) must be equal to
\be
z \int \d u\,\Omega_{(0)}\left[W^{(N-1)}\,T^{(4-N)}\right] + Q\chi_N\,,
\ee
where the BRST-trivial term $Q\chi_N$ vanishes when $W_{ij}$ is constant, i.e. $G^{(3)}=W^{(N-1)}\,T^{(4-N)}$ for  arbitrary $N$. This implies
\be
{\mathcal{M}}_N \propto \Big\langle V_\mathrm{SYM}\, z \int\d u\,\Omega_{(0)}\left[W^{(N-1)}\,T^{(4-N)}\right]\Big\rangle\,,
\ee
and thus, using (\ref{m1}),
\be
{\mathcal{M}}_N \propto \int\d^4x\int\d u\,\bD^4 D^{\prime 4}\left[W^{(1)}\, W^{(N-1)}\,T^{(4-N)}\right] = \int\d^4x\int\d u\,\bD^4 D^{\prime 4}\left[W^{(N)}\,T^{(4-N)}\right],
\ee
thus proving (\ref{mn}) in the general case.

\section{Summary}

In this paper, after reviewing the work done in \cite{fleury}, we have computed the first scattering amplitude involving pure spinor vertex operators in $\mathrm{AdS}{}_5 \times \mathrm{S}^5$.  We have verified a conjecture according to which the tree-level scattering amplitude containing a supergravity state and $N$ massless open superstring states close to the boundary of $\mathrm{AdS}{}_5$  can be written as a harmonic superspace integral involving the supergravity and super-Yang--Mills (SYM) fields. 
More precisely, we have shown that
\ba
 {\mathcal{M}}_N  & := &  \int \d\xi_1\ldots\d\xi_{N-1}\,{\Big\langle V_\mathrm{SYM}(\infty)\,V_{(N)}(\mathrm{i}\epsilon,-\mathrm{i}\epsilon)\,U_\mathrm{SYM}(\xi_1)\ldots U_\mathrm{SYM}(\xi_{N-1})\Big\rangle}_{\mathrm{D3-brane}}  \nonumber\\
 & \propto &   \int\d^4x\int\d u\,\bD^4 D^{\prime 4}\left[W^{(N)}\, T^{(4-N)}\right], \nonumber
\ea
where $V_{(N)}$ is the supergravity vertex operator defined in (\ref{vertex}) and the ``D3-brane'' subscript indicates that
the open superstring (SYM) vertex operators are located on D3-branes parallel and close to the AdS${}_5$ boundary, at some fixed value of $y^{ij}$  and $z \sim 0$.

The harmonic superspace coupling above has been known for some time \cite{howe}. Here we have shown that it can be obtained as a superstring scattering amplitude computation involving open and closed superstring vertex operators.
This can be seen as a consistency check for the vertex operator found in \cite{fleury}, as well as one more test of the AdS/CFT conjecture, in that the expected relation between supergravity and SYM was found. Future and perhaps more interesting applications would involve the computation of scattering amplitudes with closed superstring vertex operators only, which could be compared with correlation functions in the SYM side. 

\acknowledgments

We would like to thank Thiago Fleury and Renann Jusinskas for useful discussions. TA would like to thank FAPESP grant 2010/19596-2 for financial support. NB would like to thank CNPq grant 300256/94-9 and FAPESP grants 09/50639-2 and 11/11973-4 for partial financial support.


\appendix

\section{Notation and conventions}

\subsection{Two-component spinor notation}\label{twocomp}

The four-dimensional Lorentz group SO$(3,1)$ is locally isomorphic to ${\mathrm{SL(2,}\,\mathbb{C})}$, which has two distinct fundamental representations. One of them is described by a pair of complex numbers \cite{wess}
\be
\psi_\alpha = \left(\begin{array}{c} \psi_1 \cr \psi_2 \end{array}\right)\,,
\ee
with transformation law
\be
\psi_\alpha^\prime = \Lambda_\alpha^\beta \psi_\beta\,, \qquad \Lambda \in {\mathrm{SL(2,}\,\mathbb{C})}\,,
\label{rep1}
\ee
and is called $\left({1 \over 2},0\right)$ or left-handed chiral representation.

The other fundamental representation, called $\left(0,{1 \over 2}\right)$ or right-handed chiral, is obtained by complex conjugation:
\be
\bar{\psi}_\adot^\prime = \bar{\Lambda}_\adot^\bdot \bar{\psi}_\bdot \,, \qquad \bar{\Lambda}_\adot^\bdot = \overline{(\Lambda_\alpha^\beta)}\,.
\label{rep2}
\ee
The dot over the indices indicates the representation to which we refer.

The indices with and without dot are raised and lowered in the following way:
\begin{subequations}
\be
\psi^\alpha = \varepsilon^{\alpha\beta} \psi_\beta\,, \qquad \bar{\chi}^\adot = \varepsilon^{\adot\bdot}\bar{\chi}_\bdot\,;
\ee
\be
\psi_\alpha = \varepsilon_{\alpha\beta} \psi^\beta\,, \qquad \bar{\chi}_\adot = \varepsilon_{\adot\bdot}\bar{\chi}^\bdot\,,
\ee
\end{subequations}
where $\varepsilon$ is antisymmetric and has the properties
\be
\varepsilon^{12} = \varepsilon^{\dot{1}\dot{2}} = -\varepsilon_{12} = -\varepsilon_{\dot{1}\dot{2}} = 1\;\;\; \Longrightarrow \;\;\; \varepsilon_{\alpha\beta}\varepsilon^{\beta\gamma} = \delta_{\alpha}^{\gamma}\,, \qquad \varepsilon_{\adot\bdot}\varepsilon^{\bdot\dot{\gamma}} = \delta_{\adot}^{\dot{\gamma}}\,.
\ee
For spinorial derivatives, raising or lowering the indices involve an extra sign. For example, $D^\alpha_i = -\varepsilon^{\alpha\beta} D_{\beta i}$.

The convention for contraction of spinorial indices is
\be
\psi^\alpha \lambda_\alpha =: (\psi\lambda)\,, \qquad \bar{\chi}_\adot \bar{\xi}^\adot =: (\bar{\chi}\bar{\xi})\,.
\ee

In ${\mathrm{SL(2,}\,\mathbb{C})}$ notation, a four-component Dirac spinor is represented by a pair of chiral spinors:
\be
\Psi_{\mathrm{D}} = \left(\begin{array}{c} \psi_\alpha \cr \bar{\chi}^\adot \end{array}\right).
\ee
For a Majorana spinor, $\bar{\chi}_\adot = \overline{(\psi_\alpha)}$. The Dirac matrices are
\be
\Sigma^a = \left(\begin{array}{cc} 0 & (\sigma^a)_{\alpha\adot} \\ (\tilde{\sigma}^a)^{\adot\alpha} & 0 \end{array} \right),
\ee
where the matrices $\sigma^a$ ($a = 0,\ldots,3$) are defined as
\be
(\sigma^a)_{\alpha\adot} = (-{\mathbb{I}_2}, \vec{\sigma})_{\alpha\adot}\,, \qquad (\tilde{\sigma}^a)^{\adot\alpha} = \varepsilon^{\adot\bdot}\varepsilon^{\alpha\beta}(\sigma^a)_{\beta\bdot} = (-{\mathbb{I}_2}, -\vec{\sigma})^{\adot\alpha}\,,
\ee
with $\mathbb{I}_2$ the $2 \times 2$ identity matrix and $\vec{\sigma}$ the Pauli matrices
\be
\sigma^1 = 
\left(\begin{array}{cc} 0 & 1 \cr 1 & 0 \end{array}\right), \qquad \sigma^2 = \left(\begin{array}{cc} 0 & -i \cr i & 0 \end{array}\right), \qquad \sigma^3 = \left(\begin{array}{cc} 1 & 0 \cr 0 & -1 \end{array}\right),
\ee
and have the following properties:
\be
\begin{array}{l}
(\sigma^a)_{\alpha\adot}(\tilde{\sigma}_a)^{\bdot\beta} = -2\delta_\alpha^\beta \delta_\adot^\bdot\,, \qquad (\sigma_a)_{\alpha\adot}(\tilde{\sigma}^b)^{\adot\alpha} = -2\delta_a^b\,, \cr
\sigma^a\tilde{\sigma}^b = -\eta^{ab} +\sigma^{ab}\,, \qquad \tilde{\sigma}^a \sigma^b = -\eta^{ab} +\tilde{\sigma}^{ab}\,, \cr
\sigma^{ab} = -\sigma^{ba}\,, \qquad \tilde{\sigma}^{ab} = -\tilde{\sigma}^{ba}\,, \qquad (\sigma^{ab})_\alpha{}^\alpha = (\tilde{\sigma}^{ab})^\adot{}_\adot = 0\,,

\end{array}
\ee
with $\eta^{ab} = {\mathrm{diag}}(-1,1,1,1)$. These properties imply $\{\Sigma^a,\Sigma^b\}=-2\eta^{ab}\,\mathbb{I}_4$.

\subsection{Dimensional reduction}\label{dimred}

Since in the text we write expressions both in ten- and four-dimensional notation, it is important to clarify our notation and conventions. Breaking the SO($9, 1$) Lorentz symmetry to $\mathrm{SO}(3, 1) \times \mathrm{SO}(6)$ $\simeq$  $\mathrm{SO}(3, 1) \times \mathrm{SU}(4)$, an SO($9, 1$) vector $v^\mu$ ($\mu = 0, \ldots, 9$) decomposes as

\begin{equation}
v^\mu \;\longmapsto\; (v^a, v^{[ij]})\,,
\end{equation}
where $v^a$ ($a = 0,\ldots,3$) transforms under the representation {\bf 4} of SO($3, 1$) and $v^{[ij]} = -v^{[ji]}$ ($i,j = 1,\ldots,4$) transforms under the {\bf 6} of SU(4). The relation between the {\bf 6} of SU(4) and the {\bf 6} of SO(6) is given by the SO(6) Pauli matrices $(\rho_I)^{ij}=-(\rho_I)^{ji}$ ($I = 1,\ldots,6$) in the following way:
\be
v^{[ij]}=\frac{1}{2\mathrm{i}}(\rho_I)^{ij} v^{I+3}\,.
\ee
These matrices have the properties \cite{GSW}
\ba
(\rho^I)^{ij}(\rho^J)_{jk} +  (\rho^J)^{ij}(\rho^I)_{jk}   & = &   2\eta^{IJ}\delta_k^i \,, \nonumber \\
(\rho^I)_{ij}   & = &   \frac{1}{2}\varepsilon_{ijk\ell}(\rho^I)^{k\ell}\,, \\
(\rho^I)_{ij}(\rho_I)_{k\ell}   & = &   -2\varepsilon_{ijk\ell}\,, \nonumber
\ea
where $\eta^{IJ} = {\mathrm{diag}}(1,1,1,1,1,1)$ and $\varepsilon_{ijk\ell}$ is the SU(4)-invariant, totally antisymmetric tensor such that $\varepsilon_{1234}=1$. Analogously, one can define the tensor $\varepsilon^{ijk\ell}$ such that $\varepsilon^{1234}=1$. These satisfy the relation
\be
\varepsilon_{ijk\ell}\varepsilon^{k\ell mn} = 4\delta_{[i}^m \delta_{j]}^n\,.
\ee

A left-handed Majorana-Weyl spinor $\xi^\ahat$ ($\ahat = 1, \ldots, 16$) transforming under the  {\bf 16} of $\mathrm{SO}(9, 1)$ decomposes as

\be
\xi^\ahat \;\longmapsto\; (\xi^{\alpha i},\bar{\xi}_j^\adot)\,,
\ee
where we use the standard two-component notation for chiral spinors ($\alpha = 1,2\,; \adot = \dot{1},\dot{2}$) and $\xi^{\alpha i}$ (resp. $\bar{\xi}_j^\adot$) transforms under the representation {\bf 4} (resp. $\bar{\mathbf{4}}$) of SU(4). Analogous conventions apply to right-handed Majorana-Weyl spinors of $\mathrm{SO}(9, 1)$.

We also need to know how to translate the $\mathrm{SO}(9, 1)$ Pauli matrices $(\gamma^\mu)_{\ahat\bhat}$ and $(\gamma^\mu)^{\ahat\bhat}$ to the language of $\mathrm{SO}(3, 1) \times \mathrm{SU}(4)$. Based on \cite{mor}, we propose the following {\sl ansatz} for the non-vanishing components:

\ba
(\gamma^{a})_{(\alpha i) {\adot \choose j}}   & = &  \delta_i^j (\sigma^a)_{\alpha\adot} = (\gamma^{a})_{{\adot \choose j} (\alpha i)}\nonumber\\
(\gamma^{[k\ell]})_{(\alpha i) (\beta j)}   & = &  2\varepsilon_{\alpha\beta} \delta_{[i}^k \delta_{j]}^\ell \label{gammadown}\\
(\gamma^{[k\ell]})_{{\adot \choose i} {\bdot \choose j}}   & = &   \varepsilon_{\adot\bdot}\varepsilon^{ijk\ell}\nonumber
\ea
for $(\gamma^\mu)_{\ahat\bhat}$ and

\ba
(\gamma^{a})^{(\alpha i) {\adot \choose j}}   & = &  \delta_j^i (\tilde{\sigma}^a)^{\adot\alpha}  = (\gamma^{a})^{{\adot \choose j} (\alpha i)}\nonumber\\
(\gamma^{[k\ell]})^{(\alpha i) (\beta j)}   & = &  \varepsilon^{\alpha\beta}\varepsilon^{ijk\ell} \label{gammaup}\\
(\gamma^{[k\ell]})^{{\adot \choose i} {\bdot \choose j}}   & = &  2\varepsilon^{\adot\bdot} \delta_{[i}^k \delta_{j]}^\ell \nonumber
\ea
for $(\gamma^\mu)^{\ahat\bhat}$. It is straightforward to show that the above matrices satisfy the usual relation

\be
(\gamma^\mu)_{\ahat\bhat} (\gamma^\nu)^{\bhat\ghat} + (\gamma^\nu)_{\ahat\bhat} (\gamma^\mu)^{\bhat\ghat} = -2\eta^{\mu\nu}\delta_\ahat^\ghat\,,
\label{commutation}
\ee
with $\eta^{[ij][k\ell]} := \frac{1}{2} \varepsilon^{ijk\ell}$.

As an example, we show  how to obtain the dimensional reduction of the pure spinor constraints $\lambda\gamma^\mu\lambda=0$ using (\ref{gammadown}). For $\lambda\gamma^a\lambda=0$, we have
$$
\lambda^\ahat (\gamma^{a})_{\ahat\bhat} \lambda^\bhat = 0 \; \Longleftrightarrow \; \lambda^{\alpha i}(\gamma^{a})_{(\alpha i) {\adot \choose j}}\blambda^\adot_j + \blambda^\adot_j(\gamma^{a})_{{\adot \choose j} (\alpha i)}\lambda^{\alpha i} = 2\lambda^{\alpha i}(\sigma^a)_{\alpha\adot}\blambda^\adot_i = 0\,,
$$
whence
\be
\lambda^{\alpha i}\blambda^\adot_i = 0\,.
\ee
For $\lambda\gamma^{[ij]}\lambda=0$, we have
$$
\lambda^\ahat (\gamma^{[ij]})_{\ahat\bhat} \lambda^\bhat = 0 \; \Longleftrightarrow \; \lambda^{\alpha k} (\gamma^{[ij]})_{(\alpha k) (\beta \ell)}\lambda^{\beta \ell} + \blambda_k^\adot (\gamma^{[ij]})_{{\adot \choose k} {\bdot \choose \ell}} \blambda_\ell^\bdot = 2(\lambda^i\lambda^j)-\varepsilon^{ijk\ell}(\blambda_k\blambda_\ell) =0\,,
$$
whence
\be
(\lambda^i\lambda^j)=\frac{1}{2}\varepsilon^{ijk\ell}(\blambda_k\blambda_\ell)\,.
\ee

\section{Harmonic Superspace} \label{harmspace}

In this work we make use of a harmonic superspace composed by an ${\mathcal{N}} = 4$, $d=4$ Minkowski superspace and the coset space SU(4)/S(U(2)$\times$U(2)) \cite{howe,GIOS,hartwell}. In addition to the usual coordinates $x^a$, $\theta^{\alpha i}$ and $\bar{\theta}^\adot_i$, this superspace is parameterized by new variables $u$ $\in$ SU(4), called harmonic coordinates. In terms of indices, we write $u$ as ${u_{\mathcal{I}}}^i = ({u_A}^i,{\bu_\Ap}^i)$, and denote its inverse by ${u_i}^{\mathcal{I}} = ({\bu_i}^A,{u_i}^\Ap)$. The index ${\mathcal{I}}$ is transformed by the isotropy group S(U(2)$\times$U(2)) and thus splits naturally into $A$ = 1, 2 and $\Ap$ = 3, 4. The $u$'s have the following properties:
\be
\begin{array}{c}
\overline{{\bu_i}^A} = {u_A}^i\,, \qquad \overline{{u_i}^\Ap} = {\bu_\Ap}^i\,, \cr
{u_A}^i{\bu_i}^B = \delta_A^B\,, \qquad {\bu_\Ap}^i{u_i}^\Bp = \delta_\Ap^\Bp\,, \qquad {u_A}^i{u_i}^\Ap = 0\,, \cr
\varepsilon^{ijk\ell}{\bu_i}^1{\bu_j}^2{u_k}^3{u_\ell}^4 = -1\,.
\end{array}
\label{uprop}
\ee

The bars on some of the $u$'s reflect their U(1) charge, which is opposite to that of the unbarred ones. More precisely, the U(1) charge of an object is defined as the eigenvalue of the operator
\be
D_\mathrm{o} := \frac{1}{2}\left[{u_A}^i \frac{\partial}{\partial {u_A}^i} - {\bu_\Ap}^i \frac{\partial}{\partial {\bu_\Ap}^i}\right],
\ee
i.e. $\!u$ (resp. $\bu$) has U(1) charge $\frac{1}{2}$ (resp. $-\frac{1}{2}$).

The introduction of harmonic variables allows the definition of superfields which satisfy generalized chirality constraints. A superfield $\mathcal{F}$ which satisfies
\be
u_A{}^i D_{\alpha i} \mathcal{F} = u_i{}^\Ap \bD^i_\adot \mathcal{F} = 0
\ee
is said to be G-analytic, whereas a superfield which satisfies
\be
{u_A}^i \frac{\partial}{\partial {\bu_\Ap}^i} \mathcal{F} = 0
\ee
is said to be H-analytic. A superfield that is both G- and H-analytic is called an analytic superfield, for short.

In this work, the following conventions are used:
\begin{subequations}
\ba
(uu)^{ij}   & := &  \varepsilon^{AB} u_A{}^i u_B{}^j  \,,\\
(\bu\bu)^{ij}   & := &  \varepsilon^{A^\prime B^\prime} \bu_{A^\prime}{}^i \bu_{B^\prime}{}^j  \,,\\
D^4   & := &  D^\alpha_i (uu)^{ij} D^\beta_j D_{\alpha k} (uu)^{k\ell} D_{\beta \ell} \,,\\
D^{\prime 4}   & := &  D^\alpha_i (\bu\bu)^{ij} D^\beta_j D_{\alpha k} (\bu\bu)^{k\ell} D_{\beta \ell} \,,\\
\bD^4   & := &  \bD_\adot^i (\bu\bu)_{ij} \bD_\bdot^j \bD^{\adot k} (\bu\bu)_{k\ell} \bD^{\bdot \ell}  \,,\\
\bD^{\prime 4}   & := &  \bD_\adot^i (uu)_{ij} \bD_\bdot^j \bD^{\adot k} (uu)_{k\ell} \bD^{\bdot \ell}\,, \label{praque}
\ea
\end{subequations}
where $\varepsilon^{AB}$, $\varepsilon^{\Ap\Bp}$ are completely analogous to $\varepsilon^{\alpha\beta}$, $\varepsilon^{\adot\bdot}$. Using (\ref{uprop}), the following identities (among others) can be derived:
\begin{subequations}
\ba
(uu)_{ij}   & := &   \frac{1}{2} \varepsilon_{ijk\ell} (uu)^{k\ell} = \varepsilon_{\Ap\Bp}u_i{}^\Ap u_j{}^\Bp\,,\\
(\bu\bu)_{ij}   & := &   \frac{1}{2} \varepsilon_{ijk\ell} (\bu\bu)^{k\ell} = \varepsilon_{AB}\bu_i{}^A \bu_j{}^B\,,\\
(uu)_{ij}(\bu\bu)^{jk}   & = &   u_i{}^\Ap \bu_\Ap{}^k\,,\\
(uu)_{ij}(uu)^{jk}   & = &   0\,,\\
\varepsilon_{ijk\ell}   & = &   -3\, (uu)_{[ij}(\bu\bu)_{k]\ell} - 3\, (\bu\bu)_{[ij}(uu)_{k]\ell}\,.
\ea
\end{subequations}

\section{SYM equations}\label{ap.SYM}

The ${\mathcal{N}}=1$, $d=10$ super-Yang--Mills theory admits a formulation in superspace in terms of the on-shell superfields $A_\mu$ and $A_\ahat$. Defining the supercovariant derivatives as
\ba
\nabla_\mu   & := &   \partial_\mu + A_\mu\,, \nonumber\\
\nabla_\ahat   & := &   D_\ahat + A_\ahat\,, \\
D_\ahat   & := &   \frac{\partial}{\partial\theta^\ahat} + (\gamma^\mu\theta)_\ahat\partial_\mu\,, \nonumber
\ea
and the field-strength superfields as
\ba
F_{\ahat\bhat}   & := &   \{\nabla_\ahat,\nabla_\bhat\} - 2(\gamma^\mu)_{\ahat\bhat}\nabla_\mu\,, \nonumber \\
F_{\ahat\mu}   & := &   [\nabla_\ahat,\nabla_\mu]\,,\\
F_{\mu\nu}   & := &   [\nabla_\mu,\nabla_\nu]\,, \nonumber
\ea
one can show that, in the linearized theory,
\be
F_{\ahat\bhat} = 0 \; \Longleftrightarrow \; D_{(\ahat}A_{\bhat)} = (\gamma^\mu)_{\ahat\bhat}A_\mu\,,
\label{equation}
\ee
which, using (\ref{commutation}), can be written as
\be
A_\mu = -\frac{1}{16} (\gamma_\mu)^{\ahat\bhat}D_\ahat A_\bhat\,.
\label{eq2}
\ee

Substituting (\ref{eq2}) into (\ref{equation}), we get
\be
D_{(\ahat}A_{\bhat)} = -\frac{1}{16} (\gamma_\mu)_{\ahat\bhat} (\gamma^\mu)^{\ghat\dhat} D_\ghat A_\dhat\,.
\ee
Note that this equation is equivalent to $(\gamma^{\mu_1\cdots\mu_5})^{\ahat\bhat}D_\ahat A_\bhat = 0$.
For $\ahat = (\alpha i)$ and $\bhat = (\beta j)$, we have (recall $(\gamma_a)_{(\alpha i)(\beta j)}=0$)
\ba
D_{\alpha i} A_{\beta j} + D_{\beta j} A_{\alpha i}   & =  &   -\frac{1}{8}(\gamma_{[k\ell]})_{(\alpha i)(\beta j)}(\gamma^{[k\ell]})^{\ghat\dhat} D_\ghat A_\dhat \nonumber\\
   & = &   -\frac{1}{4}\varepsilon_{\alpha\beta}\left[2(D_{[i}A_{j]})+\varepsilon_{ijk\ell}(\bD^k\bar{A}^\ell)\right],
\label{prov}
\ea
whence
\be
(D_{[i}A_{j]}) =\frac{1}{2}\varepsilon_{ijk\ell}(\bD^k\bar{A}^\ell)\,.
\label{sym1}
\ee
For $\ahat = (\alpha i)$ and $\bhat = {\bdot \choose j}$, we have (recall $(\gamma_{[k\ell]})_{(\alpha i){\bdot \choose j}}=0$)
\ba
D_{\alpha i}\bar{A}^j_\bdot + \bD^j_\bdot A_{\alpha i}   &=&   -\frac{1}{8}(\gamma_a)_{(\alpha i){\bdot \choose j}}(\gamma^{a})^{\ghat\dhat} D_\ghat A_\dhat \nonumber\\
  &=&   \frac{1}{4}\delta_i^j\left(D_{\alpha k}\bar{A}^k_{\bdot} + \bar{D}^k_{\bdot}A_{\alpha k}\right).\label{sym2}
\ea

Because their $\theta=0$ components are the same (the scalars $\phi_{ij}$), we claim that
\be
W_{ij} \equiv A_{[ij]} = -\frac{1}{16}(\gamma_{[ij]})^{\ahat\bhat}D_\ahat A_\bhat\,,
\label{HAT}
\ee
where $W_{ij}$ is the Sohnius superfield of ${\mathcal{N}}=4$ SYM \cite{sohnius} and we made use of (\ref{eq2}). In four-dimensional notation, we get
\ba
W_{ij}   &=&   -\frac{1}{32}\varepsilon_{ijk\ell}\left[(\gamma^{[k\ell]})^{(\alpha p)(\beta q)}D_{\alpha p} A_{\beta q} + (\gamma^{[k\ell]})^{{\adot \choose p}{\bdot \choose q}}\bD_\adot^p\bar{A}_\bdot^q\right]\nonumber\\
  & = &   -\frac{1}{32}\varepsilon_{ijk\ell}\left[\varepsilon^{k\ell pq}(D_pA_q) + 2(\bD^{[k}\bar{A}^{\ell]})\right]\nonumber\\
  & = &   -\frac{1}{4}(D_{[i}A_{j]})\,,
\label{Wij}
\ea
where we made use of (\ref{sym1}). Note that (\ref{prov}) then implies
\be
D_{\alpha i} A_{\beta j} + D_{\beta j} A_{\alpha i} = 4\varepsilon_{\alpha\beta}W_{ij}\,.
\label{Wijaberto}
\ee

Indeed, we can show that this superfield satisfies the same constraints as the Sohnius one. First note that (\ref{sym1}) implies $(W_{ij})^\dagger = \frac{1}{2}\varepsilon^{ijk\ell}W_{k\ell}$. Then, writing $W_{jk} = -\frac{1}{8}\varepsilon_{jk\ell m}\bD^\ell_\adot\bar{A}^{\adot m}$, we have
\ba
D_{\alpha i} W_{jk}   & = &   -\frac{1}{8}\varepsilon_{jk\ell m}D_{\alpha i}\bD^\ell_\adot\bar{A}^{\adot m} \nonumber \\
    & = &   -\frac{1}{8}\varepsilon_{jk\ell m}\{D_{\alpha i},\bD^\ell_\adot\}\bar{A}^{\adot m} +\frac{1}{8}\varepsilon_{jk\ell m}\bD^\ell_\adot D_{\alpha i}\bar{A}^{\adot m} \nonumber\\
   & = &   -\frac{1}{4}\varepsilon_{jkim}(\sigma^a)_{\alpha\adot}\partial_a \bar{A}^{\adot m} + \underbrace{\frac{1}{8}\varepsilon_{jk\ell m}\bD^\ell_\adot\bD^{\adot m} A_{\alpha i}}_{=\;0}  +\frac{1}{32}\varepsilon_{jk\ell i}\bD^\ell_\adot\left(D_{\alpha p}\bar{A}^{\adot p} - \bD^{\adot p} A_{\alpha p}\right),\nonumber
\ea
where we made use of (\ref{sym2}). Therefore,
\be
D_{\alpha i} W_{jk} = D_{\alpha [i} W_{jk]}\,.
\label{sohnius1}
\ee
Moreover, it can be shown that the Hermitian conjugate of the above equation implies
\be
\bD^k_\adot W_{ij} = \frac{2}{3}\delta_{[i|}^k \bD^\ell_\adot W_{\ell|j]}\,.
\label{sohnius2}
\ee
Note that these constraints imply $(uu)^{ij}W_{ij}$ is an analytic superfield.

Studying the Bianchi identities for the field-strength superfields leads to the following (linearized) equations of motion \cite{witten}:
\begin{subequations}
\ba
(\gamma_\mu W)_\ahat   & = &   \partial_\mu A_\ahat - D_\ahat A_\mu\,,\\
D_\ahat W^\bhat   & = &   \frac{1}{2}(\gamma^{\mu\nu})_\ahat{}^\bhat F_{\mu\nu}\,,\\
D_\ahat F_{\mu\nu}   & = &   -2\,\partial_{[\mu}(\gamma_{\nu]} W)_\ahat\,,\\
(\gamma^\mu)_{\ahat\bhat} \partial_\mu W^\bhat   & = &   0\,,\\
\partial^\mu F_{\mu\nu}   & = &   0\,,
\ea
\end{subequations}
where $W^\ahat$ is the superfield whose $\theta=0$ component is the gluino $\xi^\ahat$.
These in turn imply, by dimensional reduction,
\begin{subequations}
\ba
D_{\alpha i} W_{jk}     & = &    -\varepsilon_{ijk\ell} W_\alpha^\ell\,,\\
D_{\alpha i} \bar{W}_{\bdot j}   & = &   -2\, \partial_{\alpha\bdot}W_{ij}\,,\\
D_{\alpha i} F_\beta{}^\gamma   & = &   0\,,\\
D_{\alpha i} F^\gdot{}_\bdot   & = &   -4\,\partial_{\alpha\bdot} \bar{W}^\gdot_i\,,\\
\partial_{\alpha\bdot}F^\bdot{}_\adot   & = &   0\,,
\ea
\end{subequations}
as well as their Hermitian conjugates, where $\partial_{\alpha\bdot} := (\sigma^a)_{\alpha\bdot}\partial_a$, $F_\beta{}^\gamma := (\sigma^{ab})_\beta{}^\gamma F_{ab}$ and $F^\bdot{}_\adot := (\tilde{\sigma}^{ab})^\bdot{}_\adot F_{ab}$.

\section{Dimensionally reduced expressions}

Although we do not explicitly use them in the text, we derive here, for completeness, the dimensionally reduced forms of $\Omega_{(0)\ahat\bhat}$ and $({\mathcal{T}}D^5)^{\ahat\bhat\ghat}$, as they might be useful for the reader.

\subsection{\boldmath{Reduced form of $\Omega_{(0)\ahat\bhat}$}}

Recall the definition of $\Omega_{(0)\ahat\bhat}$ given in (\ref{omega0}):
\be
\Omega_{(0)\ahat\bhat} :=  (uu)^{ij}(\gamma^\mu\widetilde{D})_\ahat(\gamma^\nu\widetilde{D})_\bhat (\widetilde{D}\gamma_{\mu\nu[ij]}\widetilde{D})\,.
\label{defOmega}
\ee
Using the {\sl ansatz} for the dimensional reduction of the $\gamma$-matrices given in (\ref{gammadown}) and (\ref{gammaup}), we find
\begin{subequations}
\be
\Omega_{(0)(\alpha i)(\beta j)} = 4\,\varepsilon_{\alpha\beta}(\bu\bu)_{ij}D^{\prime 4} -48\,(\bu\bu)_{im}(\bu\bu)_{jn}(\bu\bu)^{k\ell}D_{\alpha k}D_{\beta\ell}(\bD^m\bD^n) -4\,\varepsilon_{\alpha\beta}(\bu\bu)_{ij}\bD^4\,,
\label{first}
\ee
\be
\begin{array}{rcl}
\Omega_{(0){\adot \choose i}(\beta j)} = \Omega_{(0)(\beta j){\adot \choose i}}   & = &  
32\,(\bu\bu)^{k\ell}(\bu\bu)^{im}(\bu\bu)_{jn}(D_k D_m)D_{\beta\ell}\bD_\adot^n \\
   &  &   +\;32\,(\bu\bu)_{k\ell}(\bu\bu)_{jm}(\bu\bu)^{in}D_{\beta n}\bD_\adot^\ell(\bD^k\bD^m)\,,
\end{array}
\ee
\be
\Omega_{(0){\adot \choose i}{\bdot \choose j}} = 4\,\varepsilon_{\adot\bdot}(\bu\bu)^{ij}\bD^4
+48\,(\bu\bu)^{im}(\bu\bu)^{jn}(\bu\bu)_{k\ell}\bD_\adot^k \bD_\bdot^\ell(D_m D_n) -4\,\varepsilon_{\adot\bdot}(\bu\bu)^{ij}D^{\prime 4}\,.
\ee
\end{subequations}
Note that these coefficients can also be obtained (up to an overall factor) by imposing that $\widetilde{\Omega}_{(0)}T \propto \lambda^\ahat \lambda^\bhat\Omega_{(0)\ahat\bhat}T$ be annihilated by $\lambda^\ahat D_\ahat = \lambda^{\alpha i}D_{\alpha i} + \blambda_{\adot i}\bD^{\adot i}$ for any G-analytic superfield $T$ and using the fact that $\Omega_{(0)\ahat\bhat}$ is $\gamma$-traceless.\footnote{The $\gamma$-tracelessness condition might not be obvious from (\ref{defOmega}), but it is not difficult to show $(\gamma^\mu)^{\ahat\bhat}\Omega_{(0)\ahat\bhat} = 0$ by direct calculation using $\gamma$-matrix identities.}

\subsection{\boldmath{Reduced form of $({\mathcal{T}}D^5)^{\ahat\bhat\ghat}$}}

In principle, the dimensional reduction of ${\mathcal{T}}D^5$ (cf. (\ref{TD5})) could also be obtained directly by using the formulas (\ref{gammadown}) and (\ref{gammaup}), but that would be a very long and tedious task. Fortunately, there is an easier way of obtaining this result, as we describe in the following.

We begin by noting that, if $\Lambda_\ahat$ is the derivative of $\lambda^\ahat$, such that $\Lambda\gamma^\mu\Lambda=0$, then
\be
(\Lambda \gamma^\mu D)(\Lambda \gamma^\nu D)(\Lambda \gamma^\rho D)(D \gamma_{\mu\nu\rho} D) =\Lambda_{\ahat_1}\Lambda_{\ahat_2}\Lambda_{\ahat_3}({\mathcal{T}}D^5)^{\ahat_1\ahat_2 \ahat_3}\,.
\label{metodo}
\ee
The left-hand side of the above equation is not so difficult to compute. Up to an overall factor, we find
\ba
(\Lambda \gamma^\mu D)(\Lambda \gamma^\nu D)(\Lambda \gamma^\rho D)(D \gamma_{\mu\nu\rho} D)    & = &  
\varepsilon^{mnj\ell}(\bLambda^i\bLambda^k)(\Lambda_mD_n)(D_iD_j)(D_kD_\ell) \nonumber \\
  &  &   +\;4\, (\bLambda^j\bLambda^k)(\bLambda^\ell\bD^i)(D_iD_j)(D_kD_\ell)\nonumber\\
  &  &   +\;3\,\varepsilon^{mnj\ell}(\bLambda^i\bD^k)(\Lambda_kD_\ell)(\Lambda_mD_n)(D_iD_j) \nonumber \\
  &  &   +\;4\,(\bLambda^\ell\bLambda^k)(\Lambda_jD_\ell)(D_iD_k)(\bD^i\bD^j)\label{reducao}\\
  &  &   +\;12\,(\bLambda^i\bD^k)(\bLambda^j\bD^\ell)(\Lambda_kD_\ell)(D_iD_j) \nonumber \\
  &  &   +\;2\,\varepsilon^{mn\ell k}(\Lambda_iD_k)(\Lambda_jD_\ell)(\Lambda_mD_n)(\bD^i\bD^j)\nonumber\\
  &  &             +\;\mathrm{H.c.}\,,\nonumber
\ea
where ``H.c.'' stands for the ``Hermitian conjugate''.
One can check that the expression obtained from the above by substituting $\Lambda$ for $\lambda$ and $D$ for $\theta$ is annihilated by $\lambda^\ahat D_\ahat$.

Comparing (\ref{reducao}) and (\ref{metodo}), one can deduce the components of ${\mathcal{T}}D^5$. For example, considering the expansion of the right-hand side of (\ref{metodo}),
\be
\begin{array}{c}
\Lambda_{\alpha i}\Lambda_{\beta j}\Lambda_{\gamma k}({\mathcal{T}}D^5)^{(\alpha i)(\beta j)(\gamma k)}
+\;3\,\Lambda_{\alpha i}\Lambda_{\beta j}\bLambda_\gdot^k({\mathcal{T}}D^5)^{(\alpha i)(\beta j){\gdot \choose k}}\\
+\;3\,\Lambda_{\alpha i}\bLambda_\bdot^j\bLambda_\gdot^k({\mathcal{T}}D^5)^{(\alpha i){\bdot \choose j}{\gdot \choose k}} + \bLambda_\adot^i\bLambda_\bdot^j\bLambda_\gdot^k({\mathcal{T}}D^5)^{{\adot \choose i}{\bdot \choose j}{\gdot \choose k}}\,,
\end{array}
\label{expansao}
\ee
it is not difficult to see that there are two independent possible contributions for the term appearing in the first line of (\ref{reducao}), since $(\Lambda_i \Lambda_j) = \frac{1}{2}\varepsilon_{ijk\ell}(\bLambda^k\bLambda^\ell)$. Namely,
\be
{\left.({\mathcal{T}}D^5)^{(\alpha i)(\beta j)(\gamma k)}\right|}_{D^5\bD^0} = \kappa_1\,
\varepsilon^{\alpha\beta}\varepsilon^{ijmn}\varepsilon^{k\ell pq} D_\ell^\gamma (D_m D_p)(D_n D_q) + {}_{(\alpha i)}^{\;\;\;\nearrow} {}^{(\beta j)}_{\,\displaystyle\leftarrow} {}^{\searrow}_{(\gamma k)} 
\label{kappa1}
\ee
and
\be
{\left.({\mathcal{T}}D^5)^{{\adot \choose i}{\bdot \choose j}(\gamma k)}\right|}_{D^5\bD^0} = \kappa_2\,
\varepsilon^{\adot\bdot}\varepsilon^{k\ell pq} D_\ell^\gamma (D_i D_p)(D_j D_q)\,,
\label{kappa2}
\ee
where $\kappa_1$ and $\kappa_2$ are constants to be determined, ${\big|}_{D^5\bD^0}$ means terms containing 5 $D$'s and no $\bD$ and ${}_{(\alpha i)}^{\;\;\;\nearrow} {}^{(\beta j)}_{\,\displaystyle\leftarrow} {}^{\searrow}_{(\gamma k)}$ means cyclic permutations. Substituting (\ref{kappa1}) and (\ref{kappa2}) into (\ref{expansao}) and then comparing with (\ref{reducao}), we get
\be
3\kappa_2 - 6\kappa_1 = 1\,.
\ee
Furthermore, the $\gamma$-tracelessness condition $(\gamma^{[k\ell]})_{\ahat\bhat}({\mathcal{T}}D^5)^{\ahat\bhat\ghat}=0$ gives
\be
6\kappa_1 + 2\kappa_2 = 0\,,
\ee
whence $\kappa_1 = -\frac{1}{15}$ and $\kappa_2 = \frac{1}{5}$.

All the other components of ${\mathcal{T}}D^5$ can be obtained in a similar way and the result is listed in the following:

\begin{subequations}
\ba
({\mathcal{T}}D^5)^{(\alpha i)(\beta j)(\gamma k)}   &=&   -\frac{1}{15}\varepsilon^{\alpha\beta}\varepsilon^{ijmn}\varepsilon^{k\ell pq} D_\ell^\gamma (D_m D_p)(D_n D_q) + {}_{(\alpha i)}^{\;\;\;\nearrow} {}^{(\beta j)}_{\,\displaystyle\leftarrow} {}^{\searrow}_{(\gamma k)} \nonumber\\
  & &   -\frac{2}{3}\varepsilon^{\alpha\beta}\varepsilon^{ijpq}D^\gamma_p(D_\ell D_q)(\bD^\ell \bD^k)+ {}_{(\alpha i)}^{\;\;\;\nearrow} {}^{(\beta j)}_{\,\displaystyle\leftarrow} {}^{\searrow}_{(\gamma k)} \nonumber\\
  & &   +\;2\,D_\ell^\alpha D_m^\beta D_n^\gamma (\bD^{(i}\bD^j)\varepsilon^{k)\ell mn}\nonumber\\
  & &   -\frac{8}{15}\varepsilon^{\alpha\beta}D_\ell^\gamma(\bD^\ell \bD^{[i})(\bD^{j]} \bD^k)+ {}_{(\alpha i)}^{\;\;\;\nearrow} {}^{(\beta j)}_{\,\displaystyle\leftarrow} {}^{\searrow}_{(\gamma k)}\,,
\ea

\ba
({\mathcal{T}}D^5)^{{\adot \choose i}{\bdot \choose j}(\gamma k)}  &  = &  \frac{1}{5}
\varepsilon^{\adot\bdot}\varepsilon^{k\ell pq} D_\ell^\gamma (D_i D_p)(D_j D_q) \nonumber\\
  & &   -4\, \bD^{\adot[k|}\bD^{\bdot|\ell]} D_\ell^\gamma (D_iD_j)\nonumber\\
  & &   +\frac{8}{5}\delta^k_{(i}(D_{j)}D_\ell)D_m^\gamma \bD^{\adot [\ell|}\bD^{\bdot |m]}\nonumber\\
  & &   +\frac{2}{5}\varepsilon^{\adot\bdot}\varepsilon_{ijpq}D_\ell^\gamma (\bD^\ell \bD^p)(\bD^q \bD^k)\nonumber\\
  & &   +\, \varepsilon_{\ell mn(i} D_{j)}^\gamma \bD^{\adot \ell}\bD^{\bdot m}(\bD^n \bD^k)\nonumber\\
  & &   -\frac{1}{5}\delta_{(i}^k \varepsilon_{j)mnp} D_\ell^\gamma \bD^{\adot m} \bD^{\bdot p}(\bD^n\bD^\ell)\,,
\ea

\be
({\mathcal{T}}D^5)^{{\adot \choose i}(\beta j)(\gamma k)} = \overline{({\mathcal{T}}D^5)^{(\alpha i){\bdot \choose j}{\gdot \choose k}}}\,, \qquad ({\mathcal{T}}D^5)^{{\adot \choose i}{\bdot \choose j}{\gdot \choose k}} = \overline{({\mathcal{T}}D^5)^{(\alpha i)(\beta j)(\gamma k)}}\,.
\ee
\end{subequations}
%




\end{document}